\documentclass[
reprint,
prb,
superscriptaddress,
showpacs,
nofootinbib,
amsmath,
amssymb,
aps,
]{revtex4-2}

\usepackage{soul}

\usepackage{lmodern}
\usepackage{lineno}
\usepackage{mathtools}
\usepackage{color,soul}
\usepackage{graphicx}
\usepackage{dcolumn}
\usepackage{bm}
\usepackage{feynmp-auto}
\usepackage{amssymb,amsfonts,amsmath}
\usepackage{threeparttable}
\usepackage{textcomp}
\usepackage{float}
\usepackage{wrapfig}
\usepackage{verbatim}
\usepackage{accents}
\usepackage{fancyhdr}
\usepackage{euscript}
\usepackage{circledtext}
\usepackage{blkarray}
\usepackage{multirow}
\usepackage{tikz}
\usepackage{subfigure}
\usepackage{enumitem}
\usepackage{physics}
\usepackage[T1]{fontenc}
\usepackage{newtxtext,newtxmath}
\usetikzlibrary[topaths]
\usepackage[breaklinks=true, colorlinks, 
citecolor=blue,linkcolor=blue,urlcolor=blue]{hyperref}

\graphicspath{ {./figs/} }

\def\bef{\begin{framed}}
\def\eef{\end{framed}}
\def\be{\begin{equation}}
\def\ee{\end{equation}}
\def\ber{\begin{eqnarray}}
\def\eer{\end{eqnarray}}

\def\sigmabold{\mbox{\boldmath $\sigma$}}
\def\rv{{\bf r}}

\def\zv{{\bf \hat z}}

\def\vv{{\bf v}}

\def\kv{{\bf k}}

\def\Av{{\bf A}}
\def\Ev{{\bf E}}
\def\Pv{{\bf P}}

\def\Ev{{\bf E}}

\def\vv{{\bf v}}
\def\nn{\nonumber}

\begin{document}
\title{Nonconserved Density Accumulations in Orbital Hall Transport:\\ Insights from Linear Response Theory}
\author{Hao Sun}
\email{sun.hao@nus.edu.sg}
\affiliation{The Institute for Functional Intelligent Materials (I-FIM), National University of Singapore, 4 Science Drive 2, Singapore 117544}

\author{Alexander Kazantsev}
\affiliation{\noindent Department of Physics and Astronomy, University of Manchester, Manchester M13 9PL, UK}

\author{Alessandro Principi}
\affiliation{\noindent Department of Physics and Astronomy, University of Manchester, Manchester M13 9PL, UK}

\author{Giovanni Vignale}
\email{vgnl.g@nus.edu.sg}
\affiliation{The Institute for Functional Intelligent Materials (I-FIM), National University of Singapore, 4 Science Drive 2, Singapore 117544}
 
\date{\today}

\begin{abstract}
We present a linear response theory for stationary density accumulations in anomalous transport phenomena, such as the orbital Hall effect, where the transported density is odd under time reversal and the underlying charge is not conserved. Our framework applies to both metals and insulators, topologically trivial or nontrivial, and distinguishes between contributions from bulk and edge states, as well as undergap and dissipative currents.
In time-reversal invariant systems, we prove a microscopic reciprocity theorem showing that only dissipative currents at the Fermi level contribute to density accumulation, while undergap currents do not. In contrast, in non-time-reversal invariant systems, non-dissipative density accumulations, such as magnetoelectric polarization, can appear in both the bulk and edges. Importantly, we find that the net density accumulation does not always vanish, pointing to a global non-conservation that implies the existence of a non-vanishing integrated ``net torque'' in addition to a ``distributed torque'', which has zero spatial average.
We show that the distributed torque can be absorbed in the divergence of a redefined current that satisfies Onsager reciprocity, while the net torque must be explicitly accounted for. Finally, we apply our theory to two-dimensional models with edge terminations.
\end{abstract}

\maketitle

\section{Introduction}\label{intro}

Recent years have witnessed a revolution in the theory of electronic transport in crystalline materials~\cite{Haldane1988, Bernevig2006,Kane2005,Kane20052,Maciejko2011}. Traditional classifications of ``metals'' and ``insulators'' have given way to a more nuanced classification, recognizing that some materials can behave as insulators in the bulk while exhibiting metallic properties on their surfaces or edges. At the same time, quantum geometric features of the band structure have been shown to profoundly influence transport phenomena~\cite{Sodemann2015, Kang2019, Ma2019}.  
In particular, ``Berry curvature'' has emerged as a crucial actor, serving as a momentum space analogue of the magnetic field. Following this, many conducting materials have been found to support transverse currents (perpendicular to the electric field), which are termed ``anomalous'' because they arise in the absence of an external magnetic field~\cite{RevModPhys1986, RevModPhys2010, Gianfrate2020, Sodemann2015, Zhang2018, Kang2019, Liu2016}. 

While the anomalous Hall effect is the best-known example in this class of phenomena, our focus in this paper is on effects that entail the generation of transverse currents of {\it non-conserved} quantities, such as spin, valley, and orbital magnetic moment (OMM), all of which are electrically neutral and odd under time reversal~\cite{kai2014,Hirsch1999, Sinova2004, Go2018}. Although these currents are difficult to observe directly, they generally lead to accumulations of the corresponding densities on the surfaces or edges of the systems in which they flow. The practical importance of these ``density accumulations'' cannot be overstated since they are a direct and observable manifestation of bulk currents, and  provide a natural way to connect these currents to external devices~\cite{Kato2004,Saburo2008,Liu2018,Ruiz2022,Choi2023}. However, the process by which density accumulations are established is surprisingly complex and not completely understood, even in the well-studied case of the electric anomalous Hall effect (AHE), where the accumulation involves the conserved electric charge (this deceptively simpler case will be discussed in the concluding section).

Our primary aim in this paper is to establish a versatile linear response formalism for calculating not just the currents (as it is customary) but also the density accumulations in systems with edges or surfaces.  Interest in this problem is stimulated by a flurry of experimental papers reporting observations of orbital and valley density accumulations in transition metals and 2D layered materials~\cite{Choi2023, Arrighi2023, Sala2023}. In particular, the orbital Hall effect (OHE) is known to be stronger than the spin Hall effect (SHE) in principle, as it does not require spin-orbit coupling. While extensive theoretical work has been done on calculating orbital and spin Hall conductivities using linear response theory~\cite{Tanaka2008, Tanaka2010}, our work distinguishes itself by focusing on observable OMM density accumulations and on the complications arising from the fact that the underlying ``charge'' (OMM in this case) is not conserved.  In this context, the linear response theory is free of certain ambiguities that plague the  ``modern theory'' of semiclassical transport~\cite{Sundaram1999,Vanderbilt2018, Xiao2010, Sinitsyn_2008}. For example, modern theory views the Hall effect as the manifestation of an ``anomalous velocity'', inviting the question whether this anomalous velocity should also be included in the calculation of the orbital magnetic moments and torques. But in linear response theory, there is no ``anomalous velocity'' and the definition of the key observables is unambiguous (see Section~\ref{sec.II}).

\begin{figure}[t]
    \includegraphics[width=0.47\textwidth]{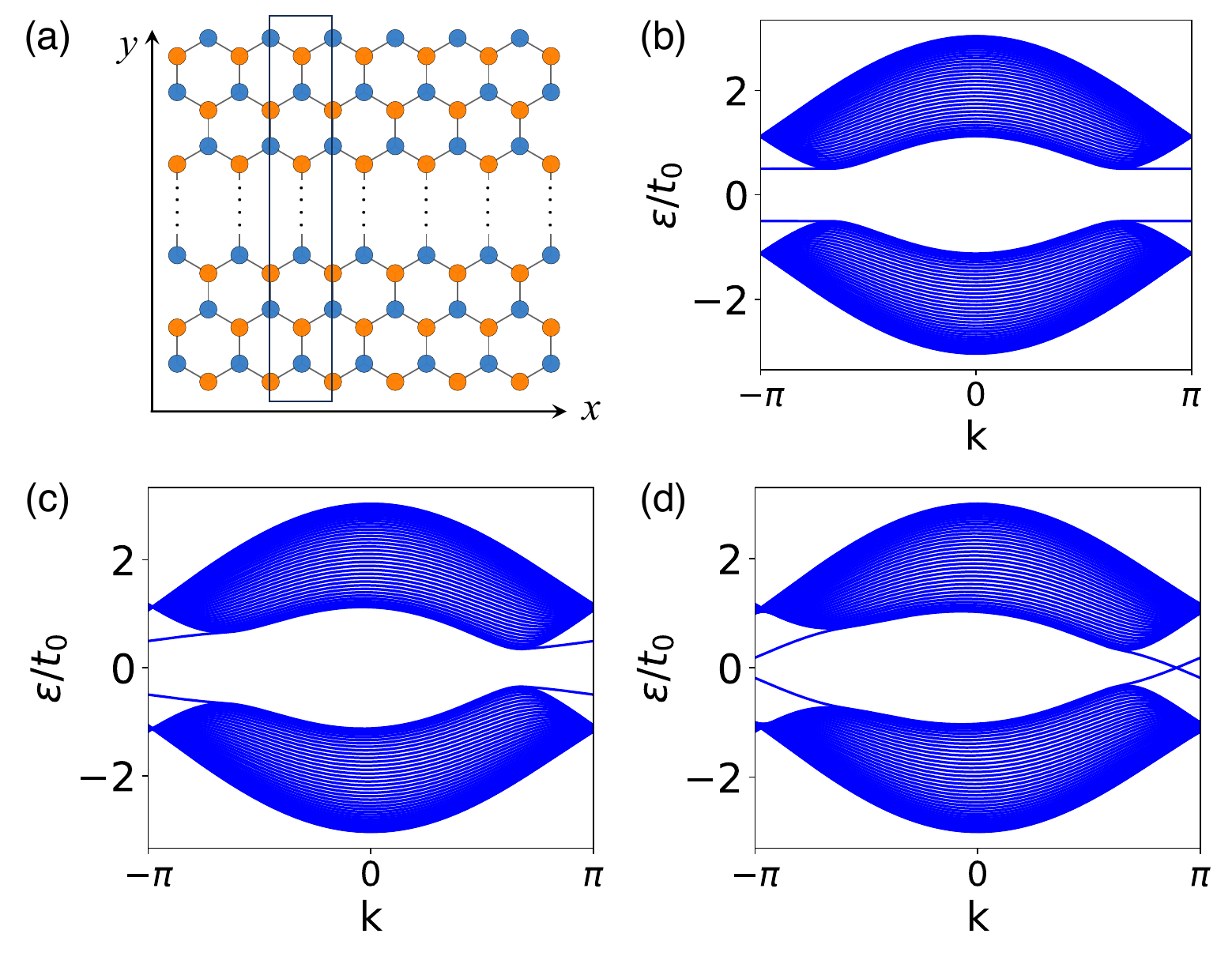}
    \caption{\label{fig1} (a) Nanoribbon model for the OMM density accumulation. (b) Band structure for graphene nanoribbon with staggered sublattice potentials. (c)  Band structure for the topologically trivial phase of the Haldane model nanoribbon with staggered sublattice potentials. (d) Band structure for the topologically nontrivial phase of the Haldane model nanoribbon with staggered sublattice potentials. In these figures, the wave vector $k$ is parallel to the edges and expressed in units of inverse lattice constant $a^{-1}$.}
\end{figure}

The model system we focus on in this paper is illustrated in Fig.~\ref{fig1}(a). A periodic system with edge or surface terminations that cut across one or two directions ($y$, $z$) is subjected to an electric field parallel to the open direction ($x$) so that an electric charge current may or may not flow depending on whether the system is a metal, a trivial insulator, or a topological insulator~\cite{Drigo2020}.  Typical band structures of our model system (practically realized, for instance, in nanoribbons of honeycomb lattices) are shown in Fig.~\ref{fig1}(b)-(d).


An important topological distinction emerges according to whether the edge states connect states on the same side of the gap or run across the gap between the bands. In the first case, illustrated in Fig.~\ref{fig1}(b)-(c), the system is said to be topologically trivial and will be considered (i) an insulator, (ii) an ``edge metal'', or (iii) a metal, depending on whether the Fermi level lies (i) within the gap without crossing any edge states,  (ii) crosses only edge states, or (iii) crosses bulk states. In the topologically nontrivial case, illustrated in Fig.~\ref{fig1}(d), the system will be considered a topological insulator (Chern insulator) if the Fermi level crosses only edge states, or a metal, if the Fermi level lies in the continuum of bulk states.

A key feature of our linear response theory is that Fermi surface contributions to the density accumulations are clearly separated from ``undergap'' contributions which arise from occupied states below the Fermi energy. This separation allows us to answer the longstanding question of whether undergap currents contribute to edge density accumulations. In TR invariant systems, we find that undergap currents do not contribute to density accumulations. The latter can only occur in metallic or edge-metallic states and are inevitably accompanied by dissipation. We refer to this result, first derived in Ref.\cite{kazantsev2024}, as the {\it ``no-dissipation no-accumulation theorem''}. 

Notice that the case of quantum spin Hall insulators requires a more careful discussion. These systems are TR invariant, yet they support spin-density accumulations and persistent edge spin currents due to spin-momentum locking. Both effects arise from the presence of edge states crossing the Fermi level in agreement with the thesis of our theorem.  However, our theorem also implies that such edge currents must be dissipative: indeed, while they are protected against elastic scattering from non-magnetic impurities, they are not protected against more general scattering processes and therefore are qualitatively different from undergap currents such as the ones that flow in the anomalous Hall effect (see further discussion in Sec.~\ref{sec.V}).

In systems with broken TR symmetry, non-dissipative edge density accumulations can occur even in topologically trivial states, i.e., in the absence of edge states crossing the Fermi level. An example of this is the bulk magnetoelectric effect, which will be discussed in the concluding section~\cite{Coh2011, Essin2010,Scaramucci2012}.

Our theoretical analysis does not explicitly cover some extrinsic effects, such as skew-scattering and side jump, in which impurity scattering acts as a {\it source} of current rather than a limiting factor. These effects, however, are implicitly contained in the exact eigenstates formulation of Sections \ref{sec.II} and \ref{sec.III}, and the only technical problem is to perform disorder averages, which is usually done with the help of diagrammatic techniques~\cite{Kovalev2009, Yang2011, Mirco2016, Hiroaki2018}. It should also be noted that these extrinsic effects occur in metals and are necessarily dissipative, so their inclusion will not change qualitative conclusions such as the no-dissipation-no-accumulation theorem for insulators.

The remainder of this paper is structured as follows. Section~\ref{sec.II} introduces the framework for studying generalized densities and current densities. We focus on electrically driven nonconserved density accumulations. The response of the generalized density and current is derived in Sec.~\ref{sec.III}, and the no-dissipation no-accumulation theorem is proved making use of the microscopic reciprocity theorem of Appendix~\ref{sec.S1}. In Sec.~\ref{sec.IV} we present the generalized continuity equation with a torque density term and discuss the potential advantages of introducing a ``proper'' current density that satisfies the macroscopic Onsager reciprocity relations~\cite{Dongwook2024,Shi2006}. Using a torque dipole density, the conventional continuity equation is recovered in Sec.~\ref{sec.V}.
In Sec.~\ref {sec.VI} we present the calculation of the OMM density response in model nanoribbon systems with TR symmetry (Sec.~\ref{sec.VI.A}) and without TR symmetry (Sec.~\ref{sec.VI.B}) -- the latter case including both trivial and nontrivial topological phases of the Haldane model.  We end with a summary and a brief discussion of the anomalous Hall effect in Sect.~\ref{sec.VII}.

\section{Basic definitions}\label{sec.II}

\subsection{Generalized densities, currents and torques}\label{sec.II.A}

Consider an extensive physical quantity (the ``charge'') represented by an operator of the form
\begin{equation}\label{}
    \begin{aligned}
        \hat{\mathcal{O}}=\sum_{i}\hat{\mathcal{O}}_{i},
    \end{aligned}
\end{equation} 
where the index $i$ runs over the particles and  $\hat{\mathcal{O}}_{i}=\hat{\mathcal{O}}\left(\hat{\mathbf{r}}_{i}, \hat{\mathbf{p}}_{i}, \hat{\mathbf{\sigma}}_{i}\cdots \right)$ is a function of the position, momentum, spin... of the $i$-th particle. This can also be written in the second-quantized form: 
\begin{equation}\label{}
    \begin{aligned}
        \hat{\mathcal{O}}=&\sum_{\alpha\beta}\mel{\alpha}{\hat{\mathcal{O}}}{\beta}\hat{c}^{\dagger}_{\alpha}\hat{c}_{\beta},
    \end{aligned}
\end{equation}
where $\ket{\alpha}$ and $\ket{\beta}$ represent  exact one-electron eigenstates. Notice that these are not necessarily Bloch states. Typical examples are the electric charge  operator ($\hat{\mathcal{O}}_{i}=-e\hat 1_i$), the spin operator ($\hat{\mathcal{O}}_{i}=\hbar\hat{\bm{\sigma}}_i/2$) and, our focus in this paper,  the OMM operator~\cite{Chang1996, Ceresoli2006, Shi2007}  $\hat{\mathcal{O}}_{i} = (-e/4) \left(\hat \rv_i \times \hat \vv_i - \hat\vv_i \times \hat \rv_i\right)$ where $\hat \rv_i$ and $\hat \vv_i$ are the position and velocity operators respectively.  Here, the velocity operator is defined as the time derivative of the position operator: $\hat\vv_i=\partial_t\hat\rv_i=i\hbar^{-1}[\hat H_E,\hat\rv_i]$, where $\hat H_E$ is the Hamiltonian of the system in the presence of the external field\footnote{Notice that the time derivative of an operator is constructed within the Heisenberg picture of  time evolution.  After doing this, we return to the Schr\"odinger picture.}:
\begin{equation}\label{}
    \begin{aligned}[b]
        \hat{H}_E=\hat{H}_0+e \sum_i \mathbf{E}\cdot\hat{ \mathbf{r}}_i.
    \end{aligned}
\end{equation}
Only the unperturbed Hamiltonian, $\hat H_0$, contributes to the commutator, so we can simply write
\be
\hat\vv_i=i\hbar^{-1}[\hat H_0,\hat\rv_i]\,.
\ee

To the operators $\hat{\mathcal{O}}_i$, we associate a one-body density operator $\hat{n}_{\mathcal{O}}(\rv)$ defined as follows 
\begin{equation}\label{}
    \begin{aligned}[b]
        \hat{n}_{\mathcal{O}}(\mathbf{r})=\sum_{i}\hat{\mathcal{O}}_{i}\star\delta(\mathbf{r}-\hat{\mathbf{r}}_i),
    \end{aligned}
\end{equation}
where the $\star$ is the symmetrized (Hermitian) product $\hat{A}\star \hat{B}=(\hat{A}\hat{B}+\hat{B}\hat{A})/2=\{\hat{A},\hat{B}\}/2$. 
This is the observable whose expectation value determines the ``density accumulation''.  Another important observable is the ``density accumulation rate'', i.e., the operator associated with the time derivative of the density, which  is obtained by taking the commutator of the density with the Hamiltonian $\hat H_E$:
\begin{equation}\label{GCE}
    \begin{aligned}[b]
        \partial_t \hat{n}_{\mathcal{O}}(\rv)=
        \sum_{i}(\partial_t\hat{\mathcal{O}}_i)\star\delta(\mathbf{r}-\hat{\mathbf{r}}_i)-\nabla_{\rv}\cdot\sum_{i}\hat{\mathcal{O}}_i\star\hat\vv_i\star\delta(\mathbf{r}-\hat{\mathbf{r}}_i)\,,
    \end{aligned}
\end{equation}
where  $\partial_t\hat{\mathcal{O}}_i=i\hbar^{-1}[\hat H_E,\hat{\mathcal{O}}_i]$.  In writing this equation, we have used the fact that the single-particle density operator obeys the equation of motion
\be
\partial_t \delta(\rv-\hat \rv_i)=-\nabla_{\rv} \cdot \hat \vv_i \star \delta(\rv-\hat \rv_i)\,.
\ee
This suggests that we define the  current density operator
\be
\hat{\mathbf{J}}_{\mathcal{O}}(\mathbf{r})=\sum_{i}\hat{\mathcal{O}}_i\star\hat \vv_i\star\delta(\mathbf{r}-\hat{\mathbf{r}}_i)
\ee
and the  torque density operator
\be\label{torque_operator}
\hat T_{\cal O} (\rv)=\sum_i\hat T_{{\cal O},i} \star\delta(\rv-\hat\rv_i)\,,
\ee
where 
\be\hat T_{{\cal O},i} \equiv i\hbar^{-1}[\hat H_E,\hat{\mathcal{O}}_i]
\ee
is the torque acting on the moment of the $i$-th particle.
Then, the expression for the density accumulation rate can be recast as a generalized continuity equation
\begin{equation}\label{MCE}
    \begin{aligned}[b]
        \partial_t\hat{n}_{\mathcal{O}}(\rv)+\nabla_{\mathbf{r}}\cdot\hat{\mathbf{J}}_{\mathbf{\mathcal{O}}}(\mathbf{r})=\hat T_{\cal O}(\rv)\,.
    \end{aligned}
\end{equation}
The divergence of the current on the left-hand side of this equation describes the flow of ``charge'' that is physically transported in or out of an infinitesimal volume centered at $\rv$, while the torque on the right-hand side describes the non-conservation of the ``charge'' within that volume. The latter vanishes when the ``charge'' is conserved: this is, of course, the case for the electric charge. But even in the familiar case of the spin density, the presence of spin-orbit interactions gives a non-zero contribution to the right-hand side of Eq.~(\ref{MCE}).  In the case of the orbital moment, the situation is much more complex, since the electric field itself breaks the conservation of the orbital moment.  In general, the total torque is the sum of two contributions: one, $i\hbar^{-1}[\hat H_0,\hat{\mathcal{O}}]$, arising from the non-conservation of the charge density, due to internal interactions with lattice and impurities  and one, $i\hbar^{-1}[e\Ev \cdot\hat \rv,\hat{\mathcal{O}}]$, arising from the action of the externally applied electric field.  

\subsection{Proper current and macroscopic Onsager reciprocity}\label{sec.II.B}

It is natural at this point to ask whether the torque term on the right-hand side of Eq.~(\ref{MCE}) can be absorbed in a redefinition of the current density so that the standard continuity equation (with zero on the right-hand side)  holds. 
This is the approach that was taken in Ref.~\cite{Shi2006} and  recently in Ref.~\cite{Xiao2021} to arrive at a ``proper'' definition of the spin current. The idea was to express the torque that appears on the right-hand side of Eq.~(\ref{MCE}) as the negative of the divergence of a torque dipole density 
\be\label{TorqueAsDivergence}
\hat T_{\cal O}(\rv)= -\nabla_{\mathbf{r}}\cdot\sum_{i}\hat{\mathbf{r}}_i\star\hat T_{{\cal O},i}\star\delta(\mathbf{r}-\hat{\mathbf{r}}_i)\,.
\ee
Moving this to the left-hand side of Eq.~(\ref{MCE}) and combining it with the divergence of the current led to a continuity equation of the standard form, i.e, 
\be
\partial_t\hat{n}_{\mathcal{O}}+\nabla_{\mathbf{r}}\cdot\hat{\mathbf{\mathcal{J}}}_{\mathbf{\mathcal{O}}}=0
\ee
with a ``proper''  current density 
\be\label{ProperCurrent}
\hat{\mathbf{\mathcal{J}}}_{\mathbf{\mathcal{O}}}=\sum_{i}\partial_t(\hat{\mathcal{O}}_i\star\hat{\mathbf{r}}_i)\star\delta(\mathbf{r}-\hat{\mathbf{r}}_i)\,.\ee

An appealing feature of this definition is that the coupling of the proper current to a spatially uniform but time-dependent vector potential $\Av(t)$ takes the form
\be\label{ProperCoupling}
-\sum_{i}\partial_t(\hat{\mathcal{O}}_i\star\hat{\mathbf{r}}_i) \cdot \Av(t)\,,
\ee
as can be seen starting from a coupling of the form $-\hat\Pv_{\cal O}\cdot \Ev(t)$, where 
$\hat\Pv_{\cal O}\equiv\sum_{i}\hat{\mathcal{O}}_i\star\hat{\mathbf{r}}_i$ is the macroscopic polarization associated with the charge $\hat {\cal O}$  and $\Ev(t)=-\partial_t\Av(t)$ is the electric field. The expression~(\ref{ProperCoupling}) is  obtained by performing a gauge transformation of the form $\exp[i\hat\Pv_{\cal O}\cdot\Av(t)]$. Crucially, this allows one to establish a{\it } macroscopic (Onsager) reciprocity relation between the response of  $\hat{\mathbf{\mathcal{J}}}_{\mathbf{\mathcal{O}}}$  to an electric field that couples to the electric polarization $\hat \Pv$ and the response of the standard electric current to an electric field that couples to the generalized polarization $\hat\Pv_{\cal O}$.  We call this reciprocity ``macroscopic'' because it applies to the response of currents to spatially uniform electric field, whereas the microscopic reciprocity relation described in Section~\ref{sec.III} and in  Appendix~\ref{sec.S1} applies to all linear response functions.

The problem with this approach is that the representation of the torque density as the divergence of a torque dipole density is possible only when the net torque density, integrated over space, is zero. This is not generally the case~\cite{Sugimoto2006,Xiao2021}. Under the action of an electric field, a net spin density or a net orbital moment may be generated. In particular, density accumulations on opposite boundaries of the finite system may have the same sign~\cite{Cysne2023}.  When this happens, thinking of the orbital moment as a quantity that is simply transported from one edge to another is no longer possible.  Furthermore, even when the spatial average of the torque is zero, Eq.~(\ref{TorqueAsDivergence}) is valid only in the limit of slowly varying density, as will be shown in detail in Appendix~\ref{sec.S4}.

Nevertheless we will see in Section~\ref{sec.IV} that it is still possible and useful (at least in the limit of slow spatial variation of the density) to absorb the torque dipole density in a ``proper'' definition of the current, along the lines of Ref.~\cite{Shi2006}, while still keeping the {\it net} torque on the right-hand side of the continuity equation. We will see that the proper current (unlike the conventional current) vanishes identically in a fully gapped time-reversal invariant insulator.

\section{Linear response theory for generalized  densities}\label{sec.III}

We start from the well-known formula~\cite{giuliani_vignale_2005} for the linear response of the expectation value of a single particle Hermitian operator $\hat A$ to a periodic external field $F$ of frequency $\omega$, which couples linearly to a single-particle Hermitian operator $\hat B$:
\be\delta A(\omega)=\chi_{A,B}(\omega)F(\omega)\,,
\ee
where  the response function $\chi_{A,B}(\omega)$ is given by
\be\label{chiAB}
\chi_{A,B}(\omega)=\sum_{\alpha\beta}\mathcal{L}^{\eta}_{\alpha\beta}(\omega)[\hat A]_{\alpha\beta}[\hat B]_{\beta\alpha}\,,
\ee
and 
\be\label{LAB}
\mathcal{L}^{\eta}_{\alpha\beta}(\omega)\equiv \frac{f_{\alpha}-f_{\beta}}{\epsilon_{\alpha}-\epsilon_{\beta}+\omega+i\eta}
\ee
is the Lindhard factor, $f_{\alpha}=\frac{1}{e^{\beta(\epsilon_{\alpha}-\mu)}+1}$ is the Fermi-Dirac average occupation of state-$\alpha$ at temperature $T$. We have introduced a compact notation for the matrix elements of any Hermitian operator $\hat A$:  
\be\label{HermiticityRelations}
[\hat A]_{\alpha\beta} \equiv \langle\alpha|\hat A|\beta\rangle =[\hat A]^*_{\beta\alpha}\,.
\ee
Equations~(\ref{chiAB}) and ~(\ref{LAB}) are valid for non-interacting systems, but our analysis is general and remains valid in fully interacting systems, as shown in Appendix~\ref{sec.S1}.

 The density response is obtained by setting $\hat A=\hat n_{\mathbf{\mathcal{O}}}(\mathbf{r})$, $\hat B=\hat \rv$, and $F(t)=e\Ev(t)$.   
Thus we have
\begin{equation}\label{den_res}
    \begin{aligned}[b]
        \delta n_{\mathcal{O}}(\mathbf{r},\omega)=e\chi_{{n_\mathcal{O}},\rv}(\mathbf{r},\omega)  \cdot\Ev(\omega),
    \end{aligned}
\end{equation}
where 
\begin{equation}\label{chino}
    \begin{aligned}[b]
        \chi_{{n_\mathcal{O}},\rv}(\mathbf{r},\omega) =\sum_{\alpha\beta}\mathcal{L}^{\eta}_{\alpha\beta}(\omega)[\hat n_{\mathbf{\mathcal{O}}}(\mathbf{r})]_{\alpha\beta}[\hat \rv]_{\beta\alpha}\,.
    \end{aligned}
\end{equation}

\subsection{Time reversal invariant systems: the no-accumulation-no-dissipation theorem}

Let us assume that the system is TR invariant. This means that for each eigenstate $|\alpha\rangle$ there is a time-reversed partner $|\tilde \alpha\rangle$ with the same energy, so that the sum over $\alpha$ and $\beta$ in Eq.~(\ref{chino}) can be replaced by a sum over $\tilde \alpha$ and $\tilde\beta$. It is shown in Appendix~\ref{sec.S1}  that TR invariance implies the {\it microscopic Onsager reciprocity relation}
\be\label{TRchi}
\chi_{A,B}(\omega)=\chi_{\tilde B,\tilde A}(\omega)\,,
\ee
where $\tilde A = \Theta \hat A^\dagger \Theta^{-1}$ is the transformation of $\hat A$ under TR.
We consider Hermitian operators of definite parity under TR, i.e., such that $\tilde A=\lambda_A \hat A$ and $\tilde B=\lambda_B \hat B$  where $\lambda=+1$ for a TR-even operator such as $\hat \rv$, and $\lambda=-1$ for a TR-odd operator such as the orbital moment density or the valley density.  Then Eq.~(\ref{TRchi}) tells us that
\begin{equation}\label{chiAB2}
    \begin{aligned}[b]
    \chi_{A,B}(\omega)&=\lambda_A\lambda_B \chi_{B,A}(\omega)\\
    &=\frac{1}{2}\left[\chi_{A,B}(\omega)+\lambda_A\lambda_B \chi_{B,A}(\omega)\right].
    \end{aligned}
\end{equation}
Making use of Eq.~(\ref{chiAB}) we readily find (see Appendix~\ref{sec.S1})
\begin{equation}\label{chiAB-TR}
    \begin{aligned}[b]
        \chi_{AB}(\omega)&=\frac{1+\lambda_{A}\lambda_{B}}{2}\mathcal{L}^{\eta}_{\alpha\beta}(\omega)\Re[A_{\alpha\beta}B_{\beta\alpha}]\\
        &+i\frac{1-\lambda_{A}\lambda_{B}}{2}\mathcal{L}^{\eta}_{\alpha\beta}(\omega)\Im[A_{\alpha\beta}B_{\beta\alpha}]\,.
    \end{aligned}
\end{equation}

If the operators $\hat A$ and $\hat B$ have opposite parities under time reversal, as is the case when $\hat n_{\mathbf{\mathcal{O}}}(\mathbf{r})$ is the orbital moment density or the valley density or the spin density, then $\lambda_A\lambda_B=-1$ and the formula~(\ref{chino}) for the linear response simplifies to 
\begin{equation}\label{chi_two_parts}
    \begin{aligned}[b]
       \chi_{n_\mathcal{O},\rv}(\mathbf{r},\omega)&=\sum_{\alpha\beta} i\mathcal{L}^{\eta}_{\alpha\beta}(\omega)\Im\{[\hat n_{\mathbf{\mathcal{O}}}(\mathbf{r})]_{\alpha\beta}[\hat \rv]_{\beta\alpha}\}\,.
    \end{aligned}
\end{equation}

We separate the real and the imaginary parts of $\mathcal{L}^{\eta}_{\alpha\beta}(\omega)$ as follows
\begin{equation}\label{LDecomposition1}
    \begin{aligned}[b]
\mathcal{L}^{\eta}_{\alpha\beta}(\omega)={P}\frac{f_{\alpha}-f_{\beta}}{\epsilon_{\alpha}-\epsilon_{\beta}+\omega}-i\pi (f_{\alpha}-f_{\beta})\delta(\epsilon_{\alpha}-\epsilon_{\beta}+\omega)\,,
    \end{aligned}
\end{equation}
where $P$ denotes the principal part: $P\frac{a}{x}\equiv\lim_{\eta\to0}\frac{ax}{x^2+\eta^2}$. In the limit of zero frequency ($\omega \to 0$) the first term reduces to $P(f_{\alpha}-f_{\beta})/(\epsilon_{\alpha}-\epsilon_{\beta})$, which is symmetric under the interchange of $\alpha$ and $\beta$.  At the same time, $\Im\{[\hat n_{\mathbf{\mathcal{O}}}]_{\alpha\beta}(\mathbf{r})[\hat \rv]_{\beta\alpha}\}$  is antisymmetric under the interchange of $\alpha$ and $\beta$ -- a fact that follows immediately from the hermiticity relations~(\ref{HermiticityRelations}).  Therefore, the first term of Eq.~(\ref{LDecomposition1}) gives a vanishing contribution when summed over $\alpha$ and $\beta$.  This leaves us with the simple result
\ber\label{DensityResponse27}
&&\chi_{n_\mathcal{O},\rv}(\mathbf{r},0)= \lim_{\omega \to 0}\nn\\&& \pi\sum_{\alpha\beta} \Im\{[\hat n_{\mathbf{\mathcal{O}}}(\mathbf{r})]_{\alpha\beta}[\hat \rv]_{\beta\alpha}\} (f_\alpha-f_\beta)\delta(\epsilon_{\alpha}-\epsilon_{\beta}+\omega)\,.\nn\\
\eer

It is evident that the response vanishes at zero temperature, unless there are pairs of states on opposite sides of the Fermi level ($f_\alpha-f_\beta \neq 0)$, separated by arbitrarily small excitation energy ($\epsilon_\beta-\epsilon_\alpha =\omega \to 0$).  The existence of these pairs of states is also a requirement for the occurrence of dissipation in a static electric field because these are the only pairs of states that can absorb energy from such a field.  We conclude that in a TR-invariant system the accumulation of a TR-odd density must necessarily vanish if there are no states at the Fermi level that can absorb energy from the electric field.  We have previously referred to this result as the ``no dissipation-no accumulation theorem''~\cite{kazantsev2024}. For a more mathematical discussion of this result, we refer the reader to Appendix~\ref{Sec1.1}. There we show that in the case of a disordered metal in the diffusive regime, under the relaxation time approximation,  Eq.~(\ref{DensityResponse27}) can be cast in the form 
\be \label{AccumulationTau}
\chi_{n_\mathcal{O},\rv}(\mathbf{r},0)= \sum_i N_i(\epsilon_F)\tau_i \left\langle \Re\{[\hat n_{\mathbf{\mathcal{O}}}(\mathbf{r})]_{\kv,\kv}[\hat \vv]_{\kv,\kv}\}\right \rangle_{FS,i}\,,
\ee
where the sum runs over the sheets of the Fermi surface (denoted by subscripts $FS,i$), the angular brackets denote an average over the $i$-th sheet of the Fermi surface, $N_i(\epsilon_F)$  and $\tau_i$ are, respectively, the partial density of states and the momentum relaxation time on the $i$-th sheet of the Fermi surface. As an illustration of this formula, the calculation of the current-induced spin polarization \cite{Garate2010,Gorini2017,Maleki2018,Johansson2021} in a disordered Rashba metal is presented in Appendix~\ref{Sec1.1}.

By contrast, consider the response of the generalized current $\mathbf{J}_{\mathcal{O}}(\mathbf{r})$ to the electric field. The current associated with a TR-odd density  is TR-even. Making use of Eq.~(\ref{chiAB-TR}) we then find (in the zero-frequency limit)

\begin{equation}\label{chijr}
    \begin{aligned}[b]
    \chi_{J_\mathcal{O},\mathbf{r}}(\mathbf{r},0)= 2 P\sum_{\alpha\beta}f_{\alpha} \frac{\Re\{[\hat{\mathbf{J}}_{\mathbf{\mathcal{O}}}(\mathbf{r})]_{\alpha\beta}[\hat \rv]_{\beta\alpha}\} }{\epsilon_{\alpha}-\epsilon_{\beta}}\,,
    \end{aligned}
\end{equation}
where 
\begin{equation}\label{}
    \begin{aligned}[b] 
        [\hat{\mathbf{J}}_{\mathbf{\mathcal{O}}}(\mathbf{r})]_{\alpha\beta}=\sum_{i}\mel{\alpha}{\hat{\mathcal{O}}_i\star(\partial_t\hat{\mathbf{r}}_i)\star\delta(\mathbf{r}-\hat{\mathbf{r}}_i)}{\beta}\,
    \end{aligned}
\end{equation}
denotes the matrix element of the generalized current, 
The current response involves contributions from all occupied states, so it can differ from zero even in an insulator: this is known as an ``undergap current''. 
In Appendix~\ref{sec.S2} we show that Eq.~\eqref{chijr} is equivalent to the familiar Berry curvature formulas  for  anomalous conductivities such as the orbital Hall conductivity.

\subsection{Broken time-reversal invariance}
Let us now discuss what happens in systems that are not TR invariant. Let us go back to the original expression~(\ref{chiAB}) of the response function of a generalized density at zero frequency, which reads:
\begin{equation}\label{}
    \begin{aligned}[b]
        \chi_{n_{\mathcal{O}},\mathbf{r}}(\mathbf{r},0)=&\sum_{\alpha\beta}\mathcal{L}^{\eta}_{\alpha\beta}(0)[\hat{n}_{\mathcal{O}}(\mathbf{r})]_{\alpha\beta}[\hat{\mathbf{r}}]_{\beta\alpha}\\
        =&\sum_{\alpha\beta}{P}\frac{f_{\alpha}-f_{\beta}}{\epsilon_{\alpha}-\epsilon_{\beta}}[\hat{n}_{\mathcal{O}}(\mathbf{r})]_{\alpha\beta}[\hat{\mathbf{r}}]_{\beta\alpha}\\
        -i\pi&\sum_{\alpha\beta}(f_{\alpha}-f_{\beta})\delta(\epsilon_{\alpha}-\epsilon_{\beta})[\hat{n}_{\mathcal{O}}(\mathbf{r})]_{\alpha\beta}[\hat{\mathbf{r}}]_{\beta\alpha}.
    \end{aligned}
\end{equation}
 Broken TR symmetry allows the first term on the right-hand side to be nonzero even when the second term vanishes.
Assuming that this is the case, we arrive at the expression
 \begin{equation}\label{undergap_chi}
    \begin{aligned}[b]
        \chi_{n_{\mathcal{O}},\mathbf{r}}(\mathbf{r},0)=2 P\sum_{\alpha\beta}f_{\alpha} \frac{\Re\left\{[\hat{n}_{\mathcal{O}}(\mathbf{r})]_{\alpha\beta}[\hat{\mathbf{r}}]_{\beta\alpha}\right\}}{\epsilon_{\alpha}-\epsilon_{\beta}}\,.
    \end{aligned}
\end{equation}
Notice the formal similarity with Eq.~(\ref{chijr}).  A fully gapped insulator with broken TR symmetry can support the accumulation of a TR-odd density, just as a TR-invariant insulator can support the {\it current} associated with a TR-odd density.  We will see concrete examples of this in the following sections.

\section{Generalized continuity equation and torque}\label{sec.IV}

We now consider the generalized continuity equation~\eqref{MCE} from the point of view of linear response theory. We start from the well-known identity\cite{giuliani_vignale_2005}
\be
-i\omega \chi_{A,B}(\omega)=-i\langle [\hat A,\hat B]\rangle +\chi_{\dot A,B}(\omega)
\ee
where the angular bracket denotes the average in the unperturbed ground state and  $\dot A = \partial_t \hat A=i\hbar^{-1}[\hat H_0,\hat A]$.
Applying this to our density response function, we get
\be
-i\omega \chi_{n_\mathcal{O},\rv}(\mathbf{r},\omega)=-i\langle [\hat n_\mathcal{O}(\rv),\hat \rv]\rangle+\chi_{\dot n_\mathcal{O},\rv}(\mathbf{r},\omega)\,.
\ee
After multiplication of this equation by $e\Ev$, on the left-hand side, we have the Fourier transform of the time derivative of the generalized density. Similarly, the first term on the right-hand side is easily recognized to be the density of ``torque'' exerted by the electric field, i.e., what we have called ``extrinsic torque density'' after Eq.~(\ref{MCE}). Because this torque is of the first order in $\Ev$, the average is taken over the equilibrium state (i.e., the state at zeroth order in $\Ev$). 

The second term on the right-hand side can further be decomposed into two parts according to Eq.~(\ref{GCE}):
\begin{equation}\label{}
    \begin{aligned}[b]
        \chi_{\dot{n}_{\mathcal{O}},\mathbf{r}}(\mathbf{r},\omega)=-\nabla_{\mathbf{r}}\cdot\chi_{j_{\mathcal{O}},\mathbf{r}}(\mathbf{r},\omega)+\chi_{T^{int}_{\mathcal{O}}, \rv}(\rv,\omega),
    \end{aligned}
\end{equation}
where $\chi_{T^{int}_{\mathcal{O}}, \rv}(\rv,\omega)$ is the response function of the ``intrinsic'' torque density operator 
\be
\hat T^{int}_{\mathcal{O}}(\rv)=\sum_i [\partial_t \hat{\mathcal{O}}_i]\star\delta(\rv-\hat\rv_i)
\ee
to the external field. Here $\partial_t \hat{\mathcal{O}}_i=i\hbar^{-1}[\hat H_0,\hat{\mathcal{O}}_i]$.  We have thus recovered the structure of the generalized continuity equation announced in Sec. II, Eq.~(\ref{MCE}).

But now we can go further and obtain an expression for the torque by computing the linear response function $\chi_{T^{int}_{\mathcal{O}},\mathbf{r}}$.  The total torque term is given by
\begin{equation}\label{TorqueFormula}
    \begin{aligned}[b]
        T_{\mathcal{O}}(\mathbf{r},\omega)=e\left[-i\langle[\hat{n}_\mathcal{O}(\mathbf{r}),\hat{\mathbf{r}}]\rangle+\chi_{T^{int}_{\mathcal{O}},\mathbf{r}}(\mathbf{r},\omega)\right]\cdot\mathbf{E}.
    \end{aligned}
\end{equation}

Let us focus on the {\it net torque} $\bar T_{\mathcal{O}}\equiv \lim_{\omega\to 0}\int T_{\mathcal{O}}(\mathbf{r},\omega) d\rv$. A non-zero value of this quantity implies that the total ``charge'' $\hat{\mathcal{O}}\equiv \sum_i\hat{\mathcal{O}}_i = \int \hat n_{\mathcal{O}}(\rv)d\rv$ is not conserved.  
The net torque associated to a TR-odd charge is even under time reversal: therefore utilizing once again Eq.~(\ref{chiAB-TR}) we find that its linear response is given by
\be\label{TorqueResponse0}
\lim_{\omega \to 0} \bar{\chi}_{T^{int}_{\mathcal{O}},\mathbf{r}}(\omega)= P\sum_{\alpha\beta}\frac{f_{\alpha}-f_{\beta}}{\epsilon_\alpha-\epsilon_\beta}[{\partial_t \hat{\mathcal{O}}}]_{\alpha\beta} [\hat \rv]_{\beta\alpha}\,.
\ee
Making use of the identity $[\partial_t \hat{\mathcal{O}}]_{\alpha\beta}=i\hbar^{-1}(\epsilon_\alpha-\epsilon_\beta)[\hat{\mathcal{O}}]_{\alpha\beta}$ and the definition of the principal part this can be rewritten as
\be\label{TorqueResponse1}
\lim_{\omega \to 0} \bar{\chi}_{T^{int}_{\mathcal{O}},\mathbf{r}}(\omega)=-\frac{1}{i\hbar}\sum_{\alpha\beta} \frac{(\epsilon_\alpha - \epsilon_\beta)^2 (f_\alpha-f_\beta)}{(\epsilon_\alpha - \epsilon_\beta)^2 + \eta^2} [\hat{\mathcal{O}}]_{\alpha\beta} [\hat{\rv}]_{\beta\alpha}\,.
\ee

Eq.~(\ref{TorqueResponse1}) is ``almost'' the negative of the first term in the square bracket of Eq.~\eqref{TorqueFormula}, which, after spatial integration, can be expressed as 
\be\label{TorqueResponse2}
\frac{1}{i\hbar}\langle[\hat{\mathcal{O}},\hat{\rv}]\rangle=\frac{1}{i\hbar}\sum_{\alpha\beta} (f_\alpha-f_\beta) [\hat{\mathcal{O}}]_{\alpha\beta} [\hat\rv]_{\beta\alpha}\,.
\ee
The difference between Eq.~(\ref{TorqueResponse1}) and Eq.~(\ref{TorqueResponse2}) arises from the seemingly innocent factor principal part factor, $\frac{(\epsilon_\alpha - \epsilon_\beta)^2}{(\epsilon_\alpha - \epsilon_\beta)^2 + \eta^2}$,  which excludes from the summation terms with $\epsilon_\alpha\simeq\epsilon_\beta$.
Combining Eqs.~(\ref{TorqueResponse1}) and (\ref{TorqueResponse2}) and using the identity  $\frac{\eta}{(\epsilon_\alpha - \epsilon_\beta)^2 + \eta^2}=\pi\delta(\epsilon_\alpha -\epsilon_\beta)$ (valid in the $\eta \to 0$ limit) we obtain
\ber\label{TorqueResponse3}
\bar T_{\mathcal{O}}&=&\frac{e\pi \eta}{\hbar}\sum_{\alpha\beta}(f_{\alpha} -f_\beta) \delta(\epsilon_\alpha-\epsilon_\beta){\rm Im}\left\{[\hat{\mathcal{O}}]_{\alpha\beta} [\hat \rv]_{\beta\alpha}\right\}\cdot\Ev\nn\\
\eer

Observe that this expression coincides, apart from the factor $\eta$, with the expression for the spatially integrated density accumulation, which is given by Eq.~(\ref{DensityResponse27}).
We can immediately conclude that the net torque is a Fermi surface property,just as the density accumulation.  We process Eq.~(\ref{TorqueResponse3}) following the same steps that led us to Eq.~(\ref{AccumulationTau}) for the density accumulation. In other words, replace $f_\alpha-f_\beta$ by $f'(\epsilon_\alpha)\delta(\epsilon_\alpha-\epsilon_\beta)$ and use $i(\epsilon_\beta-\epsilon_\alpha)[\hat \rv]_{\beta \alpha}=[\hat \vv]_{\beta\alpha}$. Then we replace the $\delta(\epsilon_\alpha-\epsilon_\beta)$ by
$\frac{(\pi\tau)^{-1}}{(\epsilon_\alpha-\epsilon_\beta)^2+\tau^{-2}}$, where $\tau^{-1}$ is the spectral width of disorder-broadened Bloch states (this is the relaxation time approximation).
The result is
\be\label{NetTorque}
\bar T_{\mathcal{O}} =\eta \left(e  N(\epsilon_F)\tau\left\langle \Re\{[\hat{\mathcal{O}}]_{\kv,\kv}[\hat \vv]_{\kv,\kv}\}\right \rangle_{FS}\cdot\Ev\right)\,,
\ee
where we have assumed, for simplicity, that the Fermi surface is a single sheet with a single relaxation time.  Thus, under the stated assumptions (i.e, in the relaxation time approximation)  the net volume torque $\bar T_{\cal O}$ in the zero-frequency limit is  the integrated density accumulation Eq.~(\ref{AccumulationTau}) multiplied by $\eta$, or, equivalently, divided by the adiabatic switching-on time $\eta^{-1}$.

It is perhaps not surprising that Eq.~(\ref{NetTorque}) goes to zero in the limit $\eta \to 0$. The reason is that the integrated ``torque'' is the time derivative of the total ``charge'' and therefore is expected to vanish in the steady state, provided that (i) a steady state exists, and (ii) the total ``charge" is described by a bounded operator. Both conditions are satisfied here, and  Eq.~(\ref{NetTorque}) is consistent with the findings of Ref.~\cite{Xiao2021}.

However, the most interesting result here is not the steady-state torque (which vanishes), but the  {\it external torque}, which, as discussed in Ref.~(\cite{kazantsev2024}) is the source of the density accumulation, acting for a time of order $\tau$ {\it before} the steady state is reached.  The external torque, which we call $\bar T^{ext}_{{\cal O}}$, can be extracted from Eq.~(\ref{NetTorque}) as
\be\label{ExternalTorque}
\bar T^{ext}_{{\cal O}} = \frac{\bar T_{\cal O}}{\eta\tau} = e  N(\epsilon_F)\left\langle \Re\{[\hat{\mathcal{O}}]_{\kv,\kv}[\hat \vv]_{\kv,\kv}\}\right \rangle_{FS}\cdot\Ev
\ee
i.e., it is the torque that we would compute from Eq.~(\ref{TorqueFormula})  if we chose the relaxation time $\tau$ to go to infinity while $\eta$ goes to zero in such a way that $\eta\tau=1$.  
In phenomenological theories of the density accumulation $\bar T_{{\cal O},ext}$ can be used as the torque that is generated by the external electric field  acting on Bloch states before impurities have had a chance to act.  Notice that Eq.~(\ref{ExternalTorque}) does not depend on the phenomenological parameter $\tau$ and can be evaluated from a purely microscopic theory.  The additional torque exerted by impurities or other mechanisms can then be included phenomenologically, for example in the relaxation time approximation.

In Appendix~\ref{sec.S3} we show that $\bar T_{\cal O}$ can be formally expressed as a Fermi-surface average of the commutator between  {\it projected} operators $\hat{\mathcal{O}}_{FS}$ and $\hat\rv_{FS}$, these being defined as the restrictions of $\hat{\mathcal{O}}$ and $\hat\rv$ respectively to the subspace of degenerate states at the Fermi level:
\be\label{Integrated_torque}
\bar T_{\mathcal{O}}=\frac{ie}{\hbar}\langle\bm{\mathcal{D}}\hat{\mathcal{O}} \rangle_{FS}\cdot\Ev\,,
\ee
where $\bm{\mathcal{D}}\hat{\mathcal{O}}$  is a short-hand for the commutator $i[\hat{\mathcal{O}}_{FS},\hat\rv_{FS}]$. The notation ${\mathcal{D}}$ is, of course, suggestive of the covariant derivative with respect to $\kv$ to which the commutator between projected operators reduces when the eigenstates of the Hamiltonian are Bloch waves (see Appendix~\ref{sec.S3}).

\section{Recovering a conventional continuity equation}\label{}
\label{sec.V}
We have so far considered only the net torque, i.e., the spatial average of the torque density. The difference between the actual torque density and its average will be referred to as ``distributed torque''.  The distributed torque integrates to zero and, therefore, can be expressed as the spatial divergence of a vector field. In the limit of slow spatial variation, this vector field is the negative of the {\it torque dipole density} 
\be\label{TorqueDipoleDensity}
D_{\mathcal{O},a}(\rv)= \sum_i \langle\hat r_{i,a}\star \hat T_{{\cal O},i} \star\delta(\rv-\hat\rv_i)\rangle\,,
\ee
where $a$ is a cartesian index (for a proof of this see Appendix~\ref{sec.S4}).
Thus we have
\be
T_{\cal O}(\rv)-\frac{1}{V}\bar T_{\cal O} = - \sum_a \nabla_a D_{{\cal O},a}(\rv),
\ee
where $V$ is the volume (or the area) of the system. As described in Sec.~\ref{sec.II.B} the distributed torque can be absorbed in a ``proper'' current density
\be\label{ProperCurrent_operator}
{\cal \hat J}_{{\cal O},a}(\rv)=\hat j_{{\cal O},a}(\rv)+\hat D_{\mathcal{ O},a}(\rv)\,,
\ee
and the generalized continuity equation takes the form
\be\label{GeneralizedContinuityEquation}
\partial_t {n}_{\mathcal{O}}(\mathbf{r},t)+\nabla_{\mathbf{r}}\cdot{\mathbf{\mathcal{J}}}_{\mathcal{O}}(\mathbf{r})=\frac{1}{V}\bar T_{\cal{O}}\,.
\ee
Notice the appearance of the spatially integrated (net) torque on the right-hand side. In practical applications $\bar T_{\cal{O}}$ would have to be approximated, for example by combining the microscopic expression~(\ref{ExternalTorque}) for the external contribution with a phenomenological expression for the internally generated relaxation.

Unlike the conventional current, the ``proper current'' is a Fermi surface property.  To see this, consider its response to an external electric field. Just as we did previously in Eqs.~(\ref{TorqueResponse0}) and (\ref{TorqueResponse2}) we can write the linear response of the  proper current as the sum of two terms: an intrinsic one
\be\label{TProperCurrent Response0}
\lim_{\omega \to 0} \bar{\chi}_{{\cal J}^{int}_{\mathcal{O},a},\mathbf{r}}(\omega)= P\sum_{\alpha\beta}\frac{f_{\alpha}-f_{\beta}}{\epsilon_\alpha-\epsilon_\beta}[\partial_t (\hat{r}_a\star \hat{\cal O})]_{\alpha\beta} [\hat \rv]_{\beta\alpha}\,,
\ee
where $\hat{r}_a \star \hat{\cal{O}}$ is a short-hand for $\sum_i\hat{r}_{i,a}\star \hat{\cal O}_i$, and an extrinsic one
\be\label{ProperCurrent Response1}
\frac{1}{i\hbar}\langle[\hat{r}_a \star \hat{\cal{O}},\rv]\rangle=\frac{1}{i\hbar}\sum_{\alpha\beta} (f_\alpha-f_\beta) [\hat{r}_a\star \hat{\cal O}]_{\alpha\beta} [\hat \rv]_{\beta\alpha}\,.
\ee

Crucially Eq.~(\ref{TProperCurrent Response0}) involves the matrix elements of a time derivative (with dynamics controlled by $H_0$) so it can be treated exactly in the same way as we treated the net torque in the early part of this section and we arrive at the Fermi surface formula
\be\label{ProperCurrent}
\bar{\mathcal{J}}_{\mathcal{O},a}=\frac{ie}{\hbar}\left\langle\bm{\mathcal{D}}\left(\hat{r}_a\star\hat{\mathcal{O}}\right)\right\rangle_{FS}\cdot\Ev\,,
\ee
where $\bm{\mathcal{D}}\left(\hat{r}_a\star\hat{\mathcal{O}}\right)$ is a short-hand for the commutator $i[\left(\hat{r}_a\star\hat{\mathcal{O}}\right)_{FS},\hat \rv_{FS}]$.
The torque and the current calculated in this section are valid in the limit of slowly varying density. Formally exact, but necessarily more complicated formulas for the torque density are reported in the Appendix~\ref{sec.S3}.

\section{Calculations of OMM density response}\label{sec.VI}
After presenting the general theory of density accumulations, we now turn to its practical application in the context of orbital magnetic responses~\cite{Salemi2022,Pezo2023,Bhowal2021}. Specifically, we focus on the accumulation of orbital magnetic moments in Dirac materials (graphene, hBN, TMDCs, etc.), both with and without TR symmetry~\cite{Ma2019,Du2019}. 
The OMM density operator is defined as
\begin{equation}\label{OMM_den_op}
    \begin{aligned}
        \hat{\mathbf{n}}_{m}(\mathbf{r})=\sum_{i}\hat{\mathbf{m}}_i\star\delta(\mathbf{r}-\hat{\mathbf{r}}_i)
    \end{aligned}
\end{equation}
where $\hat{\mathbf{m}}_i=\frac{e}{4}(\hat{\mathbf{r}}_i\times\hat{\mathbf{v}}_i-\hat{\mathbf{v}}_i\times\hat{\mathbf{r}}_i)$ is the OMM operator.\footnote{It should be borne in mind that the OMM density is {\it not} equivalent to the thermodynamic magnetization density~\cite{Shi2007}. While thermodynamic magnetization includes terms arising from the change of the quasiparticle density of states in response to a magnetic field, the OMM density arises solely from the kinetic angular momentum of the electrons. In studies of the orbital Hall effect and related transport phenomena, the experimentally measured quantity is the induced OMM density, not the thermodynamic magnetization~\cite{Choi2023}.}

As a concrete example, we consider the OMM response of the Haldane model with a staggered sublattice potential $\Delta$~\cite{Haldane1988}. The Hamiltonian for this system is given by (see Appendix~\ref{sec.S6}):
\begin{equation}\label{}
    \begin{aligned}[b]
        H &= -t_0 \sum_{\langle ij \rangle} \hat{a}^{\dagger}_i \hat{b}_j + t_2 e^{i\phi} \sum_{\langle\langle ij \rangle\rangle} \hat{a}^{\dagger}_i \hat{a}_j + t_2 e^{-i\phi} \sum_{\langle\langle ij \rangle\rangle} \hat{b}^{\dagger}_i \hat{b}_j + h.c. \\
          & \quad + \Delta \sum_i (\hat{a}^{\dagger}_i \hat{a}_i - \hat{b}^{\dagger}_i \hat{b}_i),
    \end{aligned}
\end{equation}
where $\hat{a}$ and $\hat{b}$ ($\hat{a}^{\dagger}$ and $\hat{b}^{\dagger}$) are the electron annihilation (creation) operators for the $A$ and $B$ sublattices of the honeycomb lattice. The parameter $ t_0 $ denotes the nearest-neighbor hopping amplitude between the A and B sublattices, while $ t_2 $ represents the next-nearest-neighbor hopping amplitude within the A (or B) sublattice. The phase parameter $ \phi $ is a measure of the staggered magnetic flux, breaking the TR symmetry.

For simplicity, we set $\phi = \pi/2$. The model can be solved analytically, yielding two symmetrically placed bands with energies:
\begin{equation}\label{}
    \begin{aligned}[b]
        \epsilon_{1,2}(\mathbf{k}) = \pm \sqrt{t^2_0 |\gamma_{\mathbf{k}}|^2 + \left(\Delta - 2t_2 \beta_{\mathbf{k}}\right)^2},
    \end{aligned}
\end{equation}
where $\gamma_{\mathbf{k}} = \sum_n e^{i \mathbf{k} \cdot \bm{\delta}_n}$, $\beta_{\mathbf{k}} = \sum_n \sin \left(\mathbf{k} \cdot \mathbf{l}_n\right)$, and $2t_2 \beta_{\mathbf{k}}$ is known as the Haldane mass. Here, $\bm{\delta}_n$ represents the nearest neighbor bonds and $\mathbf{l}_n$ are the next nearest neighbor bonds~\cite{Sun2023}. The nonzero value of $t_2$ paired with a non-vanishing magnetic flux $\phi$ (modulo $2\pi$)  breaks the TR symmetry, while $\Delta$ breaks the inversion symmetry. Thus, the model has sufficient flexibility to illustrate both the TR-invariant case ($t_2 = 0$) and the case of broken TR symmetry ($t_2 \neq 0$ and $\phi \neq 2n\pi$).

In the next two sections, we will consider the Haldane model in a ribbon geometry with terminations (edges) parallel to the $x$-axis, and we will use the formalism developed in the previous sections to calculate the OMM density accumulations.

First, in Section \ref{sec.VI.A} we will consider the TR-invariant case ($t_2=0$) where the Haldane model reduces to ``gapped graphene'' -- the gap arising from the broken inversion symmetry.  The band structure of the ribbon is shown in Fig.~\ref{fig1}(b), where the energy levels are plotted vs $k$ parallel to the edge.  In this case an OMM density is generated only when the Fermi level crosses the conduction or valence bands and arises entirely from dissipative intraband processes, in agreement with the ``non-dissipation no-accumulation theorem''.

Next, in Section \ref{sec.VI.B} we consider the Haldane model with $t_2>0$, $\phi=\pi/2$, which breaks TR symmetry. This model supports two distinct topological phases, determined by the competition between $\Delta$ and $2t_2 \beta_{\mathbf{k}}$. 

When $\Delta - 2t_2 \beta_{\mathbf{k}} > 0$, the model is in the topologically trivial phase with a Chern number of zero for each band~\cite{Haldane1988} and no edge states crossing the gap. The band structure of the ribbon in the topologically trivial phase is shown in Fig.~\ref{fig1}(c).  The essential difference between this and the TR invariant case is that an OMM density accumulation is generated even when the Fermi level lies in the gap: this nondissipative response arises from interband mixing and can be described as a magnetoelectric effect allowed by broken TR symmetry.   

When $\Delta - 2t_2 \beta_{\mathbf{k}} < 0$, the Haldane model enters a topologically nontrivial phase characterized by the quantum anomalous Hall effect~\cite{Cui2013}. The band structure of the ribbon is shown in Fig. 1(d). Because of the edge states crossing the gap, we now have both intraband and interband contribution to the OMM density accumulation.  
As a result, the system behaves as an orbital Chern insulator~\cite{Polshyn2020}, where anomalous nondissipative transport leads to simultaneous accumulations of both charge and OMM densities.

\subsection{The OMM density accumulation with TR symmetry}\label{sec.VI.A}
In this subsection, we perform calculations for finite ribbon structures based on Haldane's model. These ribbon structures maintain translational symmetry along the longitudinal direction ($x$-axis), while open boundary conditions are applied in the transverse direction ($y$-axis). The band structures of the ribbons are shown   in Figs.~\ref{fig1}(b)-(d) for various cases of interest.  

The general form of the OMM density response along the transverse direction is given by (see Appendix~\ref{sec.S7} for details):
\begin{equation}\label{den_res_ribbon}
    \begin{aligned}[b]
        \delta n^z_{m}(y) &= e\tau E_x\int dx \sum_{n,k} \frac{\partial f_{nk}}{\partial k} n^m_{nn}(x,y,k) \\
        &+ eE_x \int dx \sum_{nn'k} \frac{f_{nk}-f_{n'k}}{\epsilon_{nk}-\epsilon_{n'k}} n^m_{nn'}(x,y,k) A_{n'n}(k).
    \end{aligned}
\end{equation}
The right-hand side consists of two terms: the first is an intraband contribution, which is associated with dissipative processes at the Fermi surface~\cite{Go2020a, Go2020b} and involves the momentum relaxation time $\tau$. The second term comes from interband contributions, which are intrinsic to the system and do not involve $\tau$.
Both contributions are strongly influenced by the presence or absence of TR and inversion symmetry. For the intraband term, the breaking of inversion symmetry is essential, while the intrinsic contribution requires the breaking of both inversion and TR symmetries. Notably, in TR symmetric systems only the intraband term survives, consistent with the no-dissipation, no-accumulation theorem.

Let us go back to Eq.~\eqref{den_res_ribbon}, where the diagonal matrix elements of OMM density are
\begin{equation}\label{}
    \begin{aligned}[b]
        &n^{m}_{nn}(x,y,k)=\sum_{l}\Re{m_{nl}(k)\psi^{\dagger}_{lk}(x,y)\psi_{nk}(x,y)}.
    \end{aligned}
\end{equation}
and the matrix element of the OMM operator is given by (see Appendix~\ref{sec.S8}):
\begin{equation}\label{omm_ribbon}
    \begin{aligned}[b]
        &m_{nl}(k)=-\frac{e}{2\hbar}\partial_{k}[\epsilon_{nk}+\epsilon_{lk}]r^y_{nl}(k)+\\
        &\frac{e}{4i\hbar}\sum_{n'}\left(\epsilon_{lk}+\epsilon_{nk}-2\epsilon_{n'k}\right)\left[A_{nn'}(k)r^y_{n'l}(k)
        -r^y_{nn'}(k)A_{n'l}(k)\right],
    \end{aligned}
\end{equation}
where $r^y_{nl}(k)$ is the matrix element of the $y$-component of the position operator. The diagonal term $r^y_{nn}(k)$ indicates the average position of a state $\ket{nk}$ along the $y$ axis and provides a useful illustration of the positions of the edge states. 
Eq.~\eqref{omm_ribbon} exhibits a hybrid feature, combining the Berry connection in k-space with position matrix elements in real space. This distinguishes it from the conventional expression of the OMM in infinite systems~\cite{Chang1996, Xiao2007}.

\begin{figure}[t]
    \includegraphics[width=0.45\textwidth]{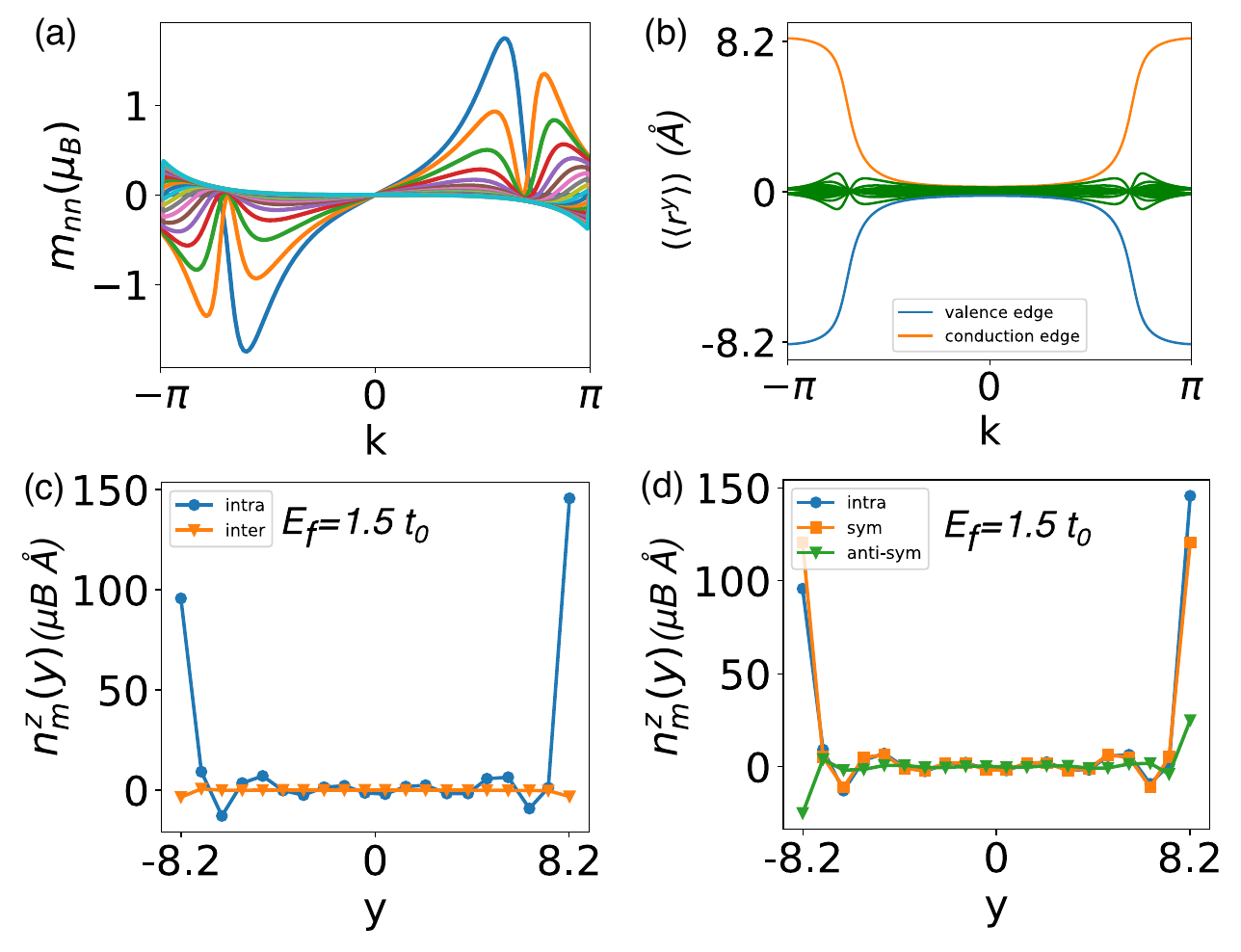} \caption{\label{fig2}  Plots of calculated properties of  TR-invariant ``gapped graphene'' nanoribbon with onsite staggered potential $\Delta=0.5t_0$ and Fermi energy  $ E_f=1.5 t_0$ [see also Fig.\ref{fig1}(b)]. (a) $k$ dependence  of the diagonal matrix element of the orbital magnetic moment  $m_{nn}(k)$. Notice that $m_{nn}(k)=-m_{nn}(-k)$, as required by TR symmetry.  (b) Plots of the centers of the ribbon states along the finite $y$ direction ($r^y_{nn}(k)$). The ribbon, schematically shown in Fig.\ref{fig1}(a)], has a width of 20 unit cells and the lattice constant is 1.42 \AA. (c) The OMM density accumulation is resolved into an intraband contribution (blue dots) and an interband  contribution (orange triangles). The interband contribution vanishes due to TR symmetry. (d) The intraband contribution to the OMM density accumulation is further resolved into  components that are  symmetric (orange squares) and antisymmetric (green triangles) with respect to the center of the ribbon ($y=0$). The existence of the symmetric component is allowed by the fact that the edge terminations break mirror symmetry in a $(z,x)$ plane passing through the center of the ribbon. (see Fig.\ref{fig1}(a)).}
\end{figure}

The OMMs, the average positions of the states along the $y$ axis, and the accumulations of the OMM density, as computed for the ribbon band structure of Fig.~\ref{fig1}(b) (gapped graphene ribbon) are presented in Fig.~\ref{fig2}. In Fig.~\ref{fig2}(a) we show the distribution of $m_{nn}(k)$ exhibiting an antisymmetric pattern in  $k$-space, a direct consequence of TR symmetry. The calculations were done for a staggered onsite potential of $\Delta=0.5t_0$ and $t_2=0.06t_0$. In Fig.~\ref{fig2}(b), we plot the centers of the ribbon states along the $y$ axis, $r^y_{nn}(k)$, which are used in calculation of the OMM density in Eq.~\eqref{omm_ribbon}. With these inputs, we calculate and plot the OMM density accumulation in Fig.~\ref{fig2}(c), where we separate, for clarity, the intraband contribution (blue) and the interband contribution (orange).  With a Fermi level crossing the bulk states at $E_f=1.5 t_0$ the interband contribution is zero, as predicted by the no-dissipation no-accumulation theorem. The accumulation arises entirely from the dissipative intraband term and its value depends on $\tau$. Further analysis, presented in Fig.~\ref{fig2}(d), shows that the induced OMM density can be decomposed into symmetric (orange squares) and antisymmetric (green triangles) components, with respect to the center of the ribbon. The symmetric component can be described as a dissipative magnetoelectric response induced by the difference between the two terminations (see Fig. Fig.\ref{fig1}(a)) --  while  the antisymmetric component originates from the orbital Hall effect~\cite{Cysne2023}.

\subsection{The OMM density accumulation with broken TR symmetry}\label{sec.VI.B}
When  TR symmetry is broken,  the OMM density response exhibits a new feature, which is especially evident when the Fermi level lies within the bulk band gap. We focus, therefore, on this case, and set $E_f=0$.  Because there is no proper Fermi surface (except for the edge states that may cross the Fermi level as will be discussed momentarily) the accumulation is dominated by the interband contribution:
\begin{equation}\label{den_res_ribbon_nondiss}
    \begin{aligned}[b]
        \delta n^z_{m}(y) = eE_x \int dx \sum_{nn'k} \frac{f_{nk} - f_{n'k}}{\epsilon_{nk} - \epsilon_{n'k}} n^m_{nn'}(x, y, k) A_{n'n}(k).
    \end{aligned}
\end{equation}
Importantly, this contribution is nondissipative and intrinsic, in sharp contrast with the Fermi surface contributions that we found in gapped graphene.

\begin{figure}[t]
    \includegraphics[width=0.45\textwidth]{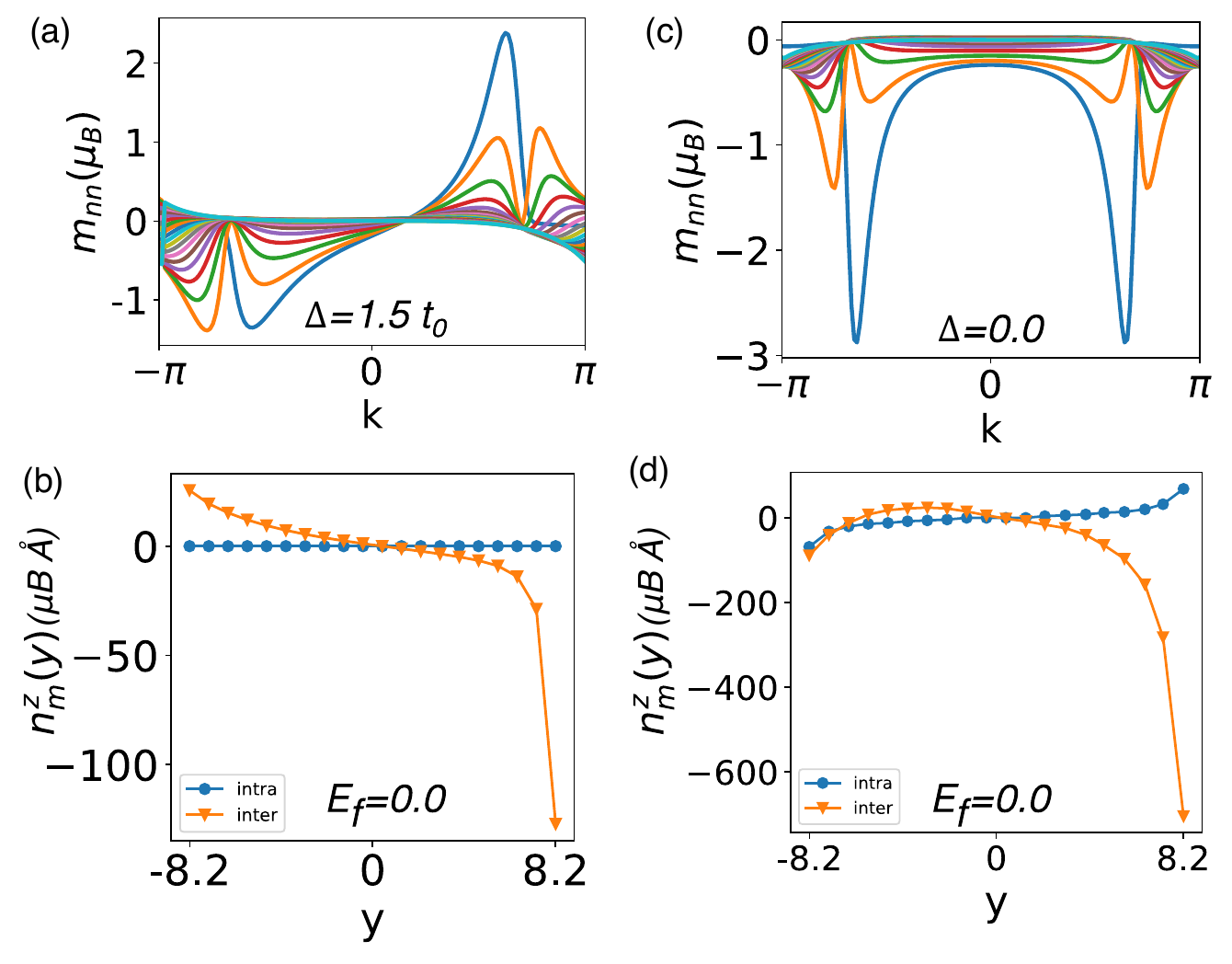}
    \caption{\label{fig3} OMM density response in Haldane model nanoribbons with flux $\phi=\frac{\pi}{2}$, and Fermi energy $E_f=0$ at the center of the gap (a) Momentum space distribution of the orbital moment in the topologically trivial phase at on-site potential $\Delta=0.5t_0$. Notice that $m_{nn}(k)\neq -m_{nn}(-k)$ due to TR symmetry breaking.  (b) The OMM density response of the trivial phase.  The intraband contribution (blue dots) is zero, while the interband contribution (orange triangles) is large and asymmetric. (c) Momentum space distribution of valence band orbital moment in the topologically nontrivial phase of the Haldane model at $\Delta=0.0$. Notice that $m_{nn}(k)=m_{nn}(-k)$ due to inversion symmetry at $\Delta=0$.  (d) The OMM density response of the nontrivial phase.  OMM density is contributed by both intraband (blue dots) and interband (orange triangles) terms. In this scenario, the intraband contribution is antisymmetric with respect to $y=0$, which leads to  opposite density accumulations at the two edges and zero net orbital magnetic response. However, the interband contribution gives rise to a large total OMM density response.}
\end{figure}

We begin with the topologically trivial phase of the Haldane model, whose ribbon band structure is shown in Fig.~\ref{fig1}(c).  In this phase, there are no edge states connecting the conduction and the valence bands. Therefore, when the Fermi level is set at midgap (zero energy) there is strictly no Fermi surface, and the intraband contribution to the OMM density accumulation vanishes as shown by the blue dots in Fig.~\ref{fig3}(b).  However, the interband contribution is quite large and spatially asymmetric (due to the breaking of inversion symmetry), as shown by the orange triangles in Fig.~\ref{fig3}(b). Crucially, the breaking of TR symmetry is essential for an orbital magnetization to appear. This can be understood by looking at the asymmetric distribution of the OMM  in momentum space, which is illustrated in Fig.~\ref{fig3}(a). The two valleys have OOMs of opposite signs but different magnitudes.  In a TR invariant system, the magnitudes would be equal so that the OMM density generated in one valley  (by virtual transitions between the conduction and valence bands)  would be exactly canceled by an opposite contribution from the other valley. In the present system the cancellation fails,  making non-dissipative orbital magnetization accumulation possible.

Further exploration of the OMM density accumulations in the topologically nontrivial phase of the Haldane model is illustrated in Fig.~\ref{fig3}(c) and (d).  The ribbon band structure is shown in Fig.~\ref{fig1}(d).  Two topologically protected edge states cross the Fermi level at midgap allowing a non-dissipative anomalous Hall current to flow in the bulk of the system.  The precise mechanism is as follows: both electric charge and orbital magnetic moment (the latter shown by the blue dots in Fig.~\ref{fig3}(d)) accumulate in the edge states.  Notice that the accumulation is antisymmetric with respect to the center of the ribbon.  The electric field generated by the charges accumulated on the edge drives the undergap Hall current.   In addition, we observe an interband response of the OMM, shown by the orange triangles in Fig.~\ref{fig3}(d), which is similar to the one observed in Fig.~\ref{fig3}(b) but significantly larger.  We attribute the larger size to the fact that the orbital moments, depicted in  Fig.~\ref{fig3}(c), have the same sign in the two valleys, leading to a much larger accumulation. This difference is evident when comparing Fig.~\ref{fig3}(c) to Fig.~\ref{fig3}(a) and is reflected in the larger magneto-electric response of the topologically nontrivial phase.  

\section{Conclusion}\label{sec.VII}

We have systematically explored the nonconserved density accumulations arising from a steady electric field in several types of conducting and nonconducting materials. The formal work has been applied to a specific phenomenon--OMM density accumulations in the orbital Hall effect. 

Firstly, we have found that in the presence of TR symmetry, the undergap contribution to density accumulation is absent. Consequently, density accumulations can only occur if there is a Fermi surface. This fact, first pointed out in Ref.~\cite{kazantsev2024} and aligning closely with recent experimental observations~\cite{Choi2023}, can be seen more directly from the fact that the ``proper current'' defined in Section~\ref{sec.IV}, absorbing the generalized torque dipole density, is a Fermi surface property.  Examining gapped graphene ribbons with TR symmetry reveals the predominant influence of intraband contributions on the accumulation of OMM density. This contribution can be further dissected into a spatially uniform component (attributed to an orbital analogue of the Edelstein effect) and an antisymmetric component, which is linked to the orbital Hall effect.

When TR symmetry is disrupted, the no-dissipation no-accumulation is no longer valid. Nondissipative accumulations can arise, for example, from interband transitions in the Haldane model.  We have illustrated the generation of a net OMM response by an electric field in a simplified Haldane model in which TR symmetry is broken. This phenomenon arises from states residing below the energy gap. It would be prohibited by the no-dissipation no-accumulation theorem in the TR symmetric scenario. 

In summary, we have provided a fresh perspective on how TR-odd nonconserved densities respond to an electric field in various types of materials. 
Remarkably, this helps us to understand the fundamental difference that exists between the quantum spin Hall effect and the quantum Hall effect.
In the quantum spin Hall effect TR symmetry requires the edge spin accumulations to go hand-in-hand with edge spin currents, the two being related by spin-momentum locking: undergap states play no role in the process.   While the spin transport at the Fermi level is almost ballistic, being protected against elastic scattering by non-magnetic impurities -- it is not truly non-dissipative, because it is not protected against other scattering mechanisms, e.g., inelastic scattering, operating at the Fermi level.  In the quantum Hall effect, edge states still play a role as hosts for the  electric charge accumulation, which produces an electric field perpendicular to the edges. However, the exactly quantized Hall current, which is perpendicular to the electric field and hence truly non-dissipative, is carried by undergap states in the incompressible bulk~\cite{Klitzing1980, Goldhaber-Gordon2015}.

\begin{acknowledgements}
    This research is supported by the Ministry of Education, Singapore, under its Research Centre of Excellence award to the Institute for Functional Intelligent Materials (I-FIM, project No. EDUNC-33-18-279-V12). A.P. and A.K. acknowledge support from the European Commission under the EU Horizon 2020 MSCA-RISE-2019 programme (project 873028 HYDROTRONICS) and from the Leverhulme Trust under the grant agreement RPG-2023-253.
\end{acknowledgements}

\bibliographystyle{apsrev4-2}
\bibliography{orb_Hall.bib}

\onecolumngrid
\clearpage
\begin{center}
\textbf{\large Supplemental Materials}
\end{center}
\setcounter{section}{0}
\setcounter{equation}{0}
\setcounter{figure}{0}
\setcounter{table}{0}
\setcounter{page}{1}
\makeatletter
\renewcommand{\theequation}{S\arabic{equation}}
\renewcommand{\thefigure}{S\arabic{figure}}
\renewcommand{\bibnumfmt}[1]{[S#1]}

\section{Microscopic reciprocity relation in time-reversal invariant systems}\label{sec.S1}
In this section we provide a general poof of Eq.~(\ref{TRchi}), which we refer to as {\it microscopic reciprocity relation} for any time-reversal invariant many-body system. 
Because we don't want to be limited to non-interacting systems we start from the general exact eigenstates (Lehmann) representation of the linear response function~\cite{giuliani_vignale_2005}
 \begin{equation}\label{res_fun}
    \begin{aligned}[b]
        \chi_{AB}(\omega) = \sum_{nm} \frac{P_m-P_n}{\omega-\omega_{nm}+i\eta}A_{mn}B_{nm}.
     \end{aligned}
\end{equation}
which describes the linear response of the Hermitian observable $\hat A$ to an external force that couples linearly to the Hermitian observable $\hat B$ (we have set $\hbar=1$). $A_{mn}=\langle m|\hat A|n\rangle$ is the matrix element of $\hat A$ between exact many-body eigenstates $|n\rangle$ and $|m\rangle$ with energies $E_n$ and $E_m$ respectively, and $\omega_{nm}=E_n-E_m$. $P_n$ and $P_m$ are the occupation probabilities of states $|n\rangle$ and $|m\rangle$ in the canonical equilibrium
ensemble at temperature $T$.
We start from the basic identity (see \cite{sakurai2020modern}, pp.~277)
\be
\langle\alpha| \hat A |\beta \rangle=\langle\tilde \beta| \hat{T} \hat A^\dagger \hat{T}^{-1} |\tilde \alpha\rangle
\ee
where $|\tilde \alpha\rangle = \hat{T}|\alpha\rangle$, $|\tilde \beta\rangle =\hat{T}|\beta\rangle$ and $\hat{T}$ is the anti-unitary time-reversal operator.

In a time-reversal invariant system for each $|n\rangle$ there is a time-reversed partner  
$|\tilde n\rangle$ with the same energy and the same occupation probability.   The sum over all $n$ is equivalent to the sum over all $\tilde n$.  Therefore we can write
\be
\chi_{AB}(\omega) = \sum_{nm} \frac{P_m-P_n}{\omega-\omega_{nm}+i\eta}\tilde B_{mn}\tilde A_{nm}= \chi_{\tilde B\tilde A}(\omega)
\ee 
where 
\be
\tilde A \equiv \hat{T} \hat A^\dagger \hat{T}^{-1}\,,~~~~\tilde B \equiv \hat{T} \hat B^\dagger \hat{T}^{-1}\,.
\ee
When dealing with operators of definite symmetry under time reversal we can further assume $\tilde A= \pm \hat A = \lambda_A \hat A$, where $\lambda_A=\pm 1$ is the signature of the operator under time reversal.  Then we have
\be
\chi_{AB}(\omega) = \lambda_A \lambda  _B\chi_{B^\dagger A^\dagger}(\omega)\,. 
\ee 
Finally, if the operators are Hermitian, we have
\be
\chi_{AB}(\omega) = \lambda_A \lambda_B  \chi_{BA}(\omega)\,. 
\ee

Now let us specialize in the case of a {\it non-interacting} system with $\hat A$ and $\hat B$ sums of one-particle-operators and $|\alpha\rangle$, $|\beta\rangle$ one-particle eigenstates.
Making use of Eqs.~(\ref{chiAB2}), (\ref{chiAB}),(\ref{LAB}) and (\ref{HermiticityRelations}) we  obtain (with summation over repeated indices implied)
\begin{equation}\label{Re_chi}
    \begin{aligned}[b]
        \Re[\chi_{AB}(\omega)]=&\frac{1}{2}\Re\left[\chi_{AB}(\omega)+\lambda_{A}\lambda_{B}\chi_{BA}(\omega)\right]\\
        =&\frac{1}{2}\Re\left[\mathcal{L}^{\eta}_{\alpha\beta}(\omega)\right]\Re[A_{\alpha\beta}B_{\beta\alpha}]-\frac{1}{2}\Im\left[\mathcal{L}^{\eta}_{\alpha\beta}(\omega)\right]\Im[A_{\alpha\beta}B_{\beta\alpha}]+\\
        &\frac{1}{2}\lambda_{A}\lambda_{B}\Re\left[\mathcal{L}^{\eta}_{\alpha\beta}(\omega)\right]\Re[B_{\alpha\beta}A_{\beta\alpha}]-\frac{1}{2}\lambda_{A}\lambda_{B}\Im\left[\mathcal{L}^{\eta}_{\alpha\beta}(\omega)\right]\Im[B_{\alpha\beta}A_{\beta\alpha}]\\
        =&\frac{1}{2}(1+\lambda_{A}\lambda_{B})\Re\left[\mathcal{L}^{\eta}_{\alpha\beta}(\omega)\right]\Re[A_{\alpha\beta}B_{\beta\alpha}]-\frac{1}{2}(1-\lambda_{A}\lambda_{B})\Im\left[\mathcal{L}^{\eta}_{\alpha\beta}(\omega)\right]\Im[A_{\alpha\beta}B_{\beta\alpha}].
    \end{aligned}
\end{equation}
where $\mathcal{L}^{\eta}_{\alpha\beta}(\omega)=\frac{f_{\alpha}-f_{\beta}}{\epsilon_{\alpha}-\epsilon_{\beta}+\omega+i\eta}$ is the Lindhard factor and 
\begin{equation}\label{}
    \begin{aligned}[b]
        \Re[\mathcal{L}^{\eta}_{\alpha\beta}(\omega)]&=(f_{\alpha}-f_{\beta})\frac{\epsilon_{\alpha}-\epsilon_{\beta}+\omega}{(\epsilon_{\alpha}-\epsilon_{\beta}+\omega)^2+\eta^2},\quad
        \Im[\mathcal{L}^{\eta}_{\alpha\beta}(\omega)]&=(f_{\alpha}-f_{\beta})\frac{-\eta}{(\epsilon_{\alpha}-\epsilon_{\beta}+\omega)^2+\eta^2}.
    \end{aligned}
\end{equation}
For the imaginary part of the response function, similarly, we have:
\begin{equation}\label{Im_chi}
    \begin{aligned}[b]
        \Im[\chi_{AB}(\omega)]=&\frac{1}{2}\Re\left[\mathcal{L}^{\eta}_{\alpha\beta}(\omega)\right]\Im[A_{\alpha\beta}B_{\beta\alpha}]+\frac{1}{2}\Im\left[\mathcal{L}^{\eta}_{\alpha\beta}(\omega)\right]\Re[A_{\alpha\beta}B_{\beta\alpha}]+\\
        &\frac{1}{2}\lambda_{A}\lambda_{B}\Re\left[\mathcal{L}^{\eta}_{\alpha\beta}(\omega)\right]\Im[B_{\alpha\beta}A_{\beta\alpha}]+\frac{1}{2}\lambda_{A}\lambda_{B}\Im\left[\mathcal{L}^{\eta}_{\alpha\beta}(\omega)\right]\Re[B_{\alpha\beta}A_{\beta\alpha}]\\
        =&\frac{1}{2}(1+\lambda_{A}\lambda_{B})\Im\left[\mathcal{L}^{\eta}_{\alpha\beta}(\omega)\right]\Re[A_{\alpha\beta}B_{\beta\alpha}]+\frac{1}{2}(1-\lambda_{A}\lambda_{B})\Re\left[\mathcal{L}^{\eta}_{\alpha\beta}(\omega)\right]\Im[A_{\alpha\beta}B_{\beta\alpha}].
    \end{aligned}
\end{equation}
Eqs.~\eqref{Re_chi} and \eqref{Im_chi} combined yield Eq.~(\ref{chiAB-TR}) of the main text:
\begin{equation}\label{}
    \begin{aligned}[b]
        \chi_{AB}(\omega)=\frac{1+\lambda_{A}\lambda_{B}}{2}\mathcal{L}^{\eta}_{\alpha\beta}(\omega)\Re[A_{\alpha\beta}B_{\beta\alpha}]+i\frac{1-\lambda_{A}\lambda_{B}}{2}\mathcal{L}^{\eta}_{\alpha\beta}(\omega)\Im[A_{\alpha\beta}B_{\beta\alpha}]\,.
    \end{aligned}
\end{equation}

\section{Generalized density accumulation in a diffusive metal}\label{Sec1.1}
In this Appendix we make use of Eq.~(\ref{DensityResponse27}) to calculate the static density accumulation.
Making use of the identities $(f_\alpha-f_\beta)\delta(\epsilon_{\alpha}-\epsilon_{\beta}+\omega)\simeq (\epsilon_\alpha-\epsilon_{\beta})f'(\epsilon_\alpha)\delta(\epsilon_{\alpha}-\epsilon_{\beta}+\omega)$ (valid in the limit $\omega\to 0$), where $f'(\epsilon_\alpha)$ is the derivative of the Fermi-Dirac distribution with respect to its own argument and $(\epsilon_{\alpha}-\epsilon_{\beta})[\hat \rv]_{\beta\alpha}=i\hbar[\hat \vv]_{\beta\alpha}$ we can write 
\begin{equation}\label{ChiN-Intermediate}
    \begin{aligned}[b]
        \chi_{n_\mathcal{O},\rv}(\mathbf{r},0)=
        \pi\hbar\sum_{\alpha\beta} f'(\epsilon_\alpha) \Re\{[\hat n_{\mathbf{\mathcal{O}}}(\mathbf{r})]_{\alpha\beta}[\hat \vv]_{\beta\alpha}\} \delta(\epsilon_{\alpha}-\epsilon_{\beta}),
    \end{aligned}
\end{equation}
where the derivative of the Fermi-Dirac distribution, $f'(\epsilon_\alpha)\simeq -\delta(\epsilon_\alpha - \epsilon_F)$ (with $\epsilon_F$ the Fermi energy) forces  $\epsilon_\alpha$ (and therefore also $\epsilon_\beta$) to be at the Fermi level. To make further progress, we need to make some assumptions about the nature of the single-particle eigenstates $|\alpha\rangle$ and $|\beta\rangle$. We work in the framework of the {\it relaxation time approximation}, namely, we assume that the eigenstates are scattering states characterized by a dominant Bloch wave vector $\kv$ with a lifetime $\tau$ due to impurity scattering.
Furthermore, we can assume that only states with the same $\kv$ are connected by the operators $\hat \rv$, $\hat \vv$, etc... This implies that only energy eigenstates separated by energies of the order of $\hbar/\tau$ are connected. While this approach is often qualitatively correct, we caution the reader that it may occasionally be spectacularly wrong, as it neglects what in diagrammatic language is termed ``vertex corrections''. When vertex corrections are important, qualitative discussion is not sufficient, and a more careful treatment of the disorder is required.

Under the assumptions of the relaxation time approximation, the sum
\be
\sum_\beta\Re\{[\hat n_{\mathbf{\mathcal{O}}}(\mathbf{r})]_{\alpha\beta}[\hat \vv]_{\beta\alpha}\}\delta(\epsilon_F-\epsilon_\beta)
\ee
can be approximated by a Lorentzian of width $1/\tau$ centered at $\epsilon_F$:
\be
 \frac{1}{\pi} \frac{\tau^{-1}}{(\epsilon_\alpha-\epsilon_F)^2+\tau^{-2}}\left\langle \Re\{\hat n_{\mathbf{\mathcal{O}}}(\mathbf{r})]_{\kv,\kv}[\hat \vv]_{\kv,\kv}\}\right \rangle_{FS} \,,
\ee
where the angular bracket denotes the average of the product of matrix elements between states at the Fermi level.
Then  Eq.~(\ref{ChiN-Intermediate}) works out to be
\be 
\chi_{n_\mathcal{O},\rv}(\mathbf{r},0)= N(\epsilon_F)\tau \left\langle \Re\{[\hat n_{\mathbf{\mathcal{O}}}(\mathbf{r})]_{\kv,\kv}[\hat \vv]_{\kv,\kv}\}\right \rangle_{FS}\,,
\ee
where $N(\epsilon_F)=\sum_\alpha\delta(\epsilon_F-\epsilon_\alpha)$ is the density of states at the Fermi level. This result is for a single sheet of the Fermi surface.  If there are multiple sheets with partial densities of states $N_i(\epsilon_F)$ and relaxation times $\tau_i$ we simply sum over them and recover Eq.~(\ref{DensityRespone27}) of the main text. 
This result diverges in the limit $\tau \to \infty$, consistent with the fact that in the absence of momentum relaxation, the electric field would drive an infinite parallel current.  Indeed $N(\epsilon_F)\tau$ (times $e^2v_F^2$) is essentially the Drude conductivity of a disordered metal in the diffusive regime. It is a measure of the shift of the Fermi surface under the action of an electric field. The fact that the induced density is proportional to this shift clearly shows that the density response is inseparable from dissipation.

As a simple illustration of this formula, consider the two-dimensional Rashba electron gas with one-particle Hamiltonian
\be\label{Rashba}
H = \frac{\hbar^2k^2}{2m}+\hbar\alpha (\zv \times \kv)\cdot \sigmabold\,
\ee
where $\alpha$ is the Rashba constant (a velocity),  $\zv$ is a unit vector perpendicular to the plane, and $\sigmabold$ is the spin. The Fermi ``surface'' (at sufficiently high electron density)  consists of two concentric circles, one at $k=k_F+m\alpha/\hbar$ (denoted by $+$) and the other at $k=k_F-m\alpha/\hbar$, (denoted by $-$), where $k_F$ is related to the two-dimensional electronic density $n$ by $n=\frac{\hbar^2k_F^2+m^2\alpha^2}{2\pi\hbar^2}$.  

We want to calculate the spin density in the $y$ direction induced by an electric field $E$ in the $x$ direction. Because the Hamiltonian~(\ref{Rashba}) is invariant under time-reversal we can use Eq.~(\ref{DensityResponse27}), which gives
\be
S_y=\frac{\hbar}{2}\left\{N_+(\epsilon_F)\tau \langle [\hat\sigma_y]_{\kv,\kv}  [\hat{{\rm v}}_x]_{\kv,\kv} \rangle_++
N_-(\epsilon_F)\tau \langle [\hat\sigma_y]_{\kv,\kv}  [\hat{{\rm v}}_x]_{\kv,\kv} \rangle_-\right\}e E\,,
\ee
where $[\hat{{\rm v}}_x]_{\kv,\kv}=\frac{1}{\hbar}\frac{\partial H}{\partial k_x}=\hbar k_x/m-\alpha[\hat\sigma_y]_{\kv,\kv}$. We have assumed that the relaxation time has the same value on both sheets of the Fermi surface.  On both Fermi circles the velocity points radially outward with a common value $v_{F+}=v_{F,-}=v_F=\hbar k_F/m$.  The spin $[\hat\sigmabold]_{\kv,\kv}$ is tangential to the circles pointing clockwise for the $+$ sheet and counterclockwise for the $-$ sheet. Finally, the density of states on the two sheets is given by
\be
N_\pm(\epsilon_F)=\frac{m}{2\pi\hbar^2}\left(1\pm \frac{\alpha}{v_F}\right)\,.
\ee
Making use of this information, it is a simple exercise to recover the well-known result for the current induced spin polarization (also known as Edelstein effect)~\cite{Edelstein1990}
\be
S_y=\frac{\hbar}{2} \frac{m}{\pi\hbar^2} \alpha \tau e E\,.
\ee

\section{Derivation of Berry curvature formulas for generalized Hall conductivities}\label{sec.S2}

Here we check that the well-known formulas connecting the Hall conductivity to the Berry curvature are recovered in our formalism. In Eq.~\eqref{res_fun} we set $\hat{A}=\hat{\mathbf{J}}_{\mathbf{\mathcal{O}}}(\mathbf{r})=\sum_{i}\hat{\mathcal{O}}_i\star\hat{\mathbf{v}}_i\star\delta(\mathbf{r}-\hat{\mathbf{r}}_i)$  (the generalized current density operator) and $\hat{B}=e\hat{\mathbf{r}}$  (the dipole operator). The response function in the DC limit ($\omega \to 0$) is given by 
\begin{equation}\label{}
    \begin{aligned}[b]
    \chi_{j_\mathcal{O},\mathbf{r}}(\mathbf{r},0)= 2 P\sum_{\alpha\beta}f_{\alpha} \frac{\Re\{[\hat{\mathbf{J}}_{\mathbf{\mathcal{O}}}]_{\alpha\beta}(\mathbf{r})[\hat \rv]_{\beta\alpha}\} }{\epsilon_{\alpha}-\epsilon_{\beta}}\,,
    \end{aligned}
\end{equation}
and the current response has the form
\begin{equation}\label{}
    \begin{aligned}[b]
        \delta J^a_{\mathcal{O}}(\mathbf{r},0)=e\sum_c \chi^{a}_{J^b_\mathcal{O},r_c}(\mathbf{r},0)E_c.
    \end{aligned}
\end{equation}
The conductivity is  given by
\begin{equation}\label{}
    \begin{aligned}[b]
        \sigma^a_{bc}(\mathbf{r})=e\chi^{a}_{J^b_\mathcal{O},r_c}(\mathbf{r},0).
    \end{aligned}
\end{equation}

In a periodic system, the state $\alpha=(n,\mathbf{k})$ is labeled by the band index $n$ and the reciprocal wave vector $\mathbf{k}$. Thus,  
\begin{equation}\label{}
    \begin{aligned}[b]
        \sigma^a_{bc}(\mathbf{r})
        =&2P\sum_{n,n',\mathbf{k}}\frac{e f_{n\mathbf{k}}}{\epsilon_{n\mathbf{k}}-\epsilon_{n'\mathbf{k}}}\Re\{[\hat{J}^{b}_{\mathcal{O}^a}]_{n\mathbf{k},n'\mathbf{k}}(\mathbf{r})A^c_{n'n}\}\\
        =&\frac{1}{2}\sum_{n\neq n',\mathbf{k}}\frac{e f_{n\mathbf{k}}}{\epsilon_{n\mathbf{k}}-\epsilon_{n'\mathbf{k}}}\Re\left[\sum_{l}\left[\{\hat{\mathcal{O}}^a,\hat{v}^b\}_{n,l}(\mathbf{k})\psi^{\dagger}_{l\mathbf{k}}\psi_{n'\mathbf{k}}(\mathbf{r})+\{\hat{\mathcal{O}}^a,\hat{v}^b\}_{l,n'}(\mathbf{k})\psi^{\dagger}_{n\mathbf{k}}\psi_{l\mathbf{k}}(\mathbf{r})\right]A^c_{n'n}\right].
    \end{aligned}
\end{equation}
Here, we have employed the expression $\mathbf{r}_{m\mathbf{k},n\mathbf{k}'}=\delta_{mn}i\frac{\partial}{\partial \mathbf{k}}\delta\left(\mathbf{k}-\mathbf{k}'\right)+\delta\left(\mathbf{k}-\mathbf{k}'\right)\mathbf{A}_{mn}(\mathbf{k})$ with $\mathbf{A}_{mn}(\mathbf{k})=\braket{u_{m\mathbf{k}}}{i\partial_{\mathbf{k}}u_{n\mathbf{k}}}$.
Thus, the global conductivity $\sigma^a_{bc}=\int \sigma^a_{bc}(\mathbf{r})d\mathbf{r}$ is  given by 
\begin{equation}\label{}
    \begin{aligned}[b]
        \sigma^a_{bc}
        =&2\sum_{n\neq n',\mathbf{k}}\frac{ef_{n\mathbf{k}}}{\epsilon_{n\mathbf{k}}-\epsilon_{n'\mathbf{k}}}\Re\left\{[\hat{\mathcal{O}}^a\star\hat{v}^b]_{n,n'}(\mathbf{k})A^c_{n'n}(\mathbf{k})\right\}\\
        =&\frac{e}{\hbar}\sum_{n,\mathbf{k}}f_{n\mathbf{k}}\sum_{n'\neq n}\frac{-2\hbar^2\Im\left\{[\hat{\mathcal{O}}^a\star\hat{v}^b]_{n,n'}(\mathbf{k})v^c_{n'n}(\mathbf{k})\right\}}{(\epsilon_{n\mathbf{k}}-\epsilon_{n'\mathbf{k}})^2}\\
        =&\frac{e}{\hbar}\sum_{n,\mathbf{k}}f_{n\mathbf{k}}\Omega^{a}_{n,bc}(\mathbf{k}),
    \end{aligned}
\end{equation}
where $A^c_{n'n}(\mathbf{k})=\frac{i\hbar v^c_{n'n}(\mathbf{k})}{\epsilon_{n\mathbf{k}}-\epsilon_{n'\mathbf{k}}}$, and $\Omega^{a}_{n,bc}(\mathbf{k})$ is the generalized Berry curvature associated with the operator $\hat{\mathcal{O}}$. Its explicit form is
\begin{equation}\label{}
    \begin{aligned}[b]
        \Omega^{a}_{n,bc}(\mathbf{k})
        =\sum_{n'\neq n}\frac{-2\hbar^2\Im\left\{[\hat{\mathcal{O}}^a\star\hat{v}^b]_{n,n'}(\mathbf{k})v^c_{n'n}(\mathbf{k})\right\}}{(\epsilon_{n\mathbf{k}}-\epsilon_{n'\mathbf{k}})^2}.
    \end{aligned}
\end{equation}

\section{The torque density}\label{sec.S3}
In this section, we provide a brief derivation of Eq.~\eqref{Integrated_torque} from the main text. We start from Eq.~\eqref{TorqueFormula}, which we restate here for convenience:
\begin{equation}\label{torque_density_0}
    \begin{aligned}[b]
        T_{\mathcal{O}}(\mathbf{r},\omega) = e \left[ \chi_{T_{\mathcal{O}},\mathbf{r}}(\omega,\mathbf{r}) + \frac{1}{i\hbar}\sum_i \langle [\hat{n}_\mathcal{O}, \hat{\mathbf{r}}_i] \rangle_0 \right] \cdot \mathbf{E},
    \end{aligned}
\end{equation}
The net torque we consider in the main text is given by 
\begin{equation}\label{integrated_t}
    \begin{aligned}[b]
        \bar{T}_{\mathcal{O}}=\lim_{\omega\to 0}\int T_{\mathcal{O}}(\mathbf{r},\omega)d\mathbf{r}=e\mathbf{E}\cdot\lim_{\omega\to 0}\bar{\chi}_{T_{\mathcal{O}},\mathbf{r}}(\omega)+\frac{e}{i\hbar}\int \sum_i \langle [\hat{n}_\mathcal{O}, \hat{\mathbf{r}}_i] \rangle_0 \cdot \mathbf{E}d\mathbf{r}
    \end{aligned}
\end{equation}
where the first term on the right-hand side is given as
\begin{equation}\label{torque_density_1}
    \begin{aligned}[b]
        \lim_{\omega\to 0}\bar{\chi}_{T_{\mathcal{O}},\mathbf{r}}(\omega)=P\sum_{\alpha\beta}\frac{f_{\alpha}-f_{\beta}}{\epsilon_{\alpha}-\epsilon_{\beta}}[\hat{\mathbf{r}}]_{\beta\alpha}
        \int T^{\mathcal{O}}_{\alpha\beta}(\mathbf{r})d\mathbf{r}\,.
    \end{aligned}
\end{equation}
The matrix element of the torque density operator is given by
\begin{equation}\label{}
    \begin{aligned}[b]
        T^{\mathcal{O}}_{\alpha\beta}(\mathbf{r})&=\mel{\alpha}{\hat{T}_{\mathcal{O}}(\mathbf{r})}{\beta}=\frac{1}{i\hbar}\sum_{i}\mel{\alpha}{[\hat{\mathcal{O}}_i,H_0]\star\delta(\mathbf{r}-\hat{\mathbf{r}}_i)}{\beta}.
    \end{aligned}
\end{equation}
Making use of the identity
\begin{equation}\label{}
    \begin{aligned}[b]
        \int T^{\mathcal{O}}_{\alpha\beta}(\mathbf{r}) d\mathbf{r}=\frac{1}{i\hbar}\sum_{i}\mel{\alpha}{[\hat{\mathcal{O}}_i,H_0]}{\beta}=\frac{1}{i\hbar}(\epsilon_{\beta}-\epsilon_{\alpha})\mathcal{O}_{\alpha\beta}\,,
    \end{aligned}    
\end{equation}
we obtain 
\begin{equation}\label{integrated_torque_1}
    \begin{aligned}[b]
        \lim_{\omega\to 0}\bar{\chi}_{T_{\mathcal{O}},\mathbf{r}}(\omega)=-\frac{1}{i\hbar}\sum_{\alpha\beta}P\left(f_{\alpha}-f_{\beta}\right)
        [\hat{\mathcal{O}}]_{\alpha\beta}[\hat{\mathbf{r}}]_{\beta\alpha},
    \end{aligned}
\end{equation}
where $[\hat{\mathcal{O}}]_{\alpha\beta}[\hat{\mathbf{r}}]_{\beta\alpha}$ is shorthand for $\sum_i [\hat{\mathcal{O}}_i]_{\alpha\beta}[\hat{\mathbf{r}}_i]_{\beta\alpha}$. This is  Eq.~\eqref{TorqueResponse1} of the main text. 

Now let us calculate the integrated commutator term in Eq.~\eqref{integrated_t}, we have
\begin{equation}\label{integrated_torque_2}
    \begin{aligned}[b]
    \frac{1}{i\hbar}\int \sum_i \langle [\hat{n}_\mathcal{O}, \hat{\mathbf{r}}_i] \rangle_0 d\mathbf{r}=\frac{1}{i\hbar}\sum_i \sum_{\alpha}[\hat{\mathcal{O}}_i, \hat{\mathbf{r}}_i]_{\alpha\alpha}f_{\alpha}=\frac{1}{i\hbar} \sum_{\alpha\beta}\left(f_{\alpha}-f_{\beta}\right)[\hat{\mathcal{O}}]_{\alpha\beta}[\hat{\mathbf{r}}]_{\beta\alpha}.
    \end{aligned}
\end{equation}
We thus recover Eq.~\eqref{TorqueResponse2} in the main text.

The combination of Eqs.~(\ref{integrated_torque_1}) and \eqref{integrated_torque_2} leads to massive cancelation, after which only the terms only with $\epsilon_{\alpha}\simeq\epsilon_{\beta}$, as given by Eq.~\eqref{} in the main text. In order to express the result more compactly, we separate the position operator $\hat{\mathbf{r}}$ into two parts, intraband $\hat{\mathbf{R}}$ and interband $\hat{\mathbf{X}}$. After employing the commutators 
\begin{equation}\label{}
    \begin{aligned}[b][\hat{\mathcal{O}},\hat{\mathbf{R}}]_{\alpha,\beta}&=-i\frac{\partial \mathcal{O}_{\alpha\beta}}{\partial \mathbf{k}}-\mathcal{O}_{\alpha\beta}(\mathbf{A}_{\alpha\alpha}-\mathbf{A}_{\beta\beta}),\\
    [\hat{\mathcal{O}},\hat{\mathbf{X}}]_{\alpha,\beta}&=\sum_{\gamma}(1-\delta_{\gamma\beta})\mathcal{O}_{\alpha\gamma}\mathbf{A}_{\gamma\beta}-\sum_{\gamma}(1-\delta_{\gamma\alpha})\mathbf{A}_{\alpha\gamma}\mathcal{O}_{\gamma\beta},
    \end{aligned}
\end{equation} 
we finally obtain
\begin{equation}\label{integrated_total_torque}
    \begin{aligned}[b]
        \bar{T}_{\mathcal{O}}=&e\mathbf{E}\cdot\sum_{\alpha}\left\{-\frac{1}{\hbar}\frac{\partial \mathcal{O}_{\alpha\alpha}}{\partial \mathbf{k}}-\frac{i}{\hbar}[\hat{\mathcal{O}},\hat{\mathbf{A}}]_{\alpha\alpha}+\frac{i}{\hbar}\sum_{\beta\neq\alpha}\left[\mathcal{O}_{\alpha\beta}A_{\beta\alpha}-\mathcal{O}_{\beta\alpha}A_{\alpha\beta}\right]\right\}f_{\alpha}\\
        =&-\frac{e}{\hbar}\sum_{\alpha}\mathbf{E}\cdot\frac{\partial \mathcal{O}_{\alpha\alpha}}{\partial \mathbf{k}}f_{\alpha}\\
        =&\frac{ie}{\hbar}\langle\mathcal{D}\hat{\mathcal{O}}\rangle_{FS}\cdot\mathbf{E},
    \end{aligned}    
\end{equation}
where we have introduced the  notation 
\be
\langle\mathcal{D}\hat{\mathcal{O}}\rangle_{FS}=i\sum_{\alpha}\frac{\partial \mathcal{O}_{\alpha\alpha}}{\partial \mathbf{k}}f_{\alpha}=-i\sum_{\alpha}\mathcal{O}_{\alpha\alpha}\frac{\partial f_{\alpha}}{\partial \mathbf{k}}\,,
\ee
which is the explicit form of the commutator  $i[\hat{\mathcal{O}}_{FS},\hat{\mathbf{r}}_{FS}]$ in Bloch system. Using integration by parts, we can see this is the Fermi surface contribution of the net torque. 

We now turn our attention to deriving the proper current formula, as presented in Eq.~\eqref{ProperCurrent} of Sec.~\ref{sec.IV}. Compared to Eq.~\eqref{torque_operator}, the proper current density operator shares a similar structure, expressed as
\begin{equation}
\hat{\mathbf{\mathcal{J}}}_{\mathbf{\mathcal{O}}} = \sum_{i} \partial_t (\hat{\mathcal{O}}_i \star \hat{\mathbf{r}}_i) \star \delta(\mathbf{r} - \hat{\mathbf{r}}_i),
\end{equation}
Thus, deriving the proper current response becomes as straightforward as replacing $\hat{\mathcal{O}}$ with $ \hat{\mathcal{O}} \star \hat{\mathbf{r}} $ in Eq.~\eqref{integrated_total_torque}. This yields
\begin{equation}\label{integrated_propercurrent}
    \bar{\mathcal{J}}_{\mathcal{O}} = \frac{e}{\hbar} \sum_{\alpha} \mathbf{E} \cdot \frac{\partial f_{\alpha}}{\partial \mathbf{k}} [\hat{\mathcal{O}} \star \hat{\mathbf{r}}]_{\alpha \alpha}
    =\frac{ie}{\hbar}\langle\mathcal{D}(\hat{\mathcal{O}} \star \hat{\mathbf{r}})\rangle_{FS}\cdot\mathbf{E},
\end{equation}
thus, we recover the Eq.~\eqref{ProperCurrent}, which clearly shows a Fermi surface contribution.

\section{Expression of torque density in the limit of slow spatial
variation}\label{sec.S4}
In this section, we show how the non-uniform component of the torque density  can be expressed, in the limit of slow spatial variation,  as  the divergence of the torque dipole density. We know that $\hat{T}_{\mathcal{O}}(\mathbf{r})=\sum_{i}(\partial_t \hat{\mathcal{O}}_i)\star\delta(\mathbf{r}-\hat{\mathbf{r}}_i)$, which in $q-$space, with Fourier transformation $\hat{T}_{\mathcal{O}}(\mathbf{q})=\int d\mathbf{r}e^{-i\mathbf{q}\cdot\mathbf{r}}\hat{T}_{\mathcal{O}}(\mathbf{r})$, is given by
\begin{equation}\label{}
    \begin{aligned}[b]
        \hat{T}_{\mathcal{O}}(\mathbf{q})=\sum_{i}(\partial_t \hat{\mathcal{O}}_i)\star e^{-i\mathbf{q}\cdot\hat{\mathbf{r}}_i}.
    \end{aligned}
\end{equation}
For $\mathbf{q}=0$, $\hat{T}_{\mathcal{O}}(\mathbf{0})$ gives the uniform component of the torque, $\hat{\bar{T}}_{\mathcal{O}}$. For $\mathbf{q}\neq0$ and for slow spatial variation ($qa\ll 1$) we expand 
\begin{equation}\label{}
    \begin{aligned}[b]
        \hat{T}_{\mathcal{O}}(\mathbf{q})\approx\mathbf{q}\cdot\sum_{i}\partial_{\mathbf{q}}\left[(\partial_t \hat{\mathcal{O}}_i)\star e^{-i\mathbf{q}\cdot\hat{\mathbf{r}}_i}\right]=-i\mathbf{q}\cdot\sum_{i}\hat{\mathbf{r}}_i\star(\partial_t \hat{\mathcal{O}}_i)\star e^{-i\mathbf{q}\cdot\hat{\mathbf{r}}_i}=-i\mathbf{q}\cdot \hat{\mathbf{D}}_{\mathcal{O}}(\mathbf{q}).
    \end{aligned}
\end{equation}
where $\hat{\mathbf{D}}_{\mathcal{O}}(\mathbf{q})=\sum_{i}\hat{\mathbf{r}}_i\star(\partial_t \hat{\mathcal{O}}_i)\star e^{-i\mathbf{q}\cdot\hat{\mathbf{r}}_i}$ is the torque dipole density operator. By transforming this expression to  real space, we have 
\begin{equation}\label{}
    \begin{aligned}[b]
        \hat{\mathbf{D}}_{\mathcal{O}}(\mathbf{r})=\int d\mathbf{q}\sum_{i}\hat{\mathbf{r}}_i\star(\partial_t \hat{\mathcal{O}}_i)\star e^{i\mathbf{q}\cdot(\mathbf{r}-\hat{\mathbf{r}}_i)}=\sum_{i}\hat{\mathbf{r}}_i\star(\partial_t \hat{\mathcal{O}}_i)\star \delta(\mathbf{r}-\hat{\mathbf{r}}_i).
    \end{aligned}
\end{equation}
This is the right operator form of Eq.~\eqref{TorqueDipoleDensity} in the main text. And the torque density operator now is rewritten as 
\begin{equation}\label{}
    \begin{aligned}[b]
        \hat{T}_{\mathcal{O}}(\mathbf{r})\approx\frac{1}{V}\hat{\bar{T}}_{\mathcal{O}}-\nabla_{\mathbf{r}}\cdot\hat{\mathbf{D}}_{\mathcal{O}}(\mathbf{r}).
    \end{aligned}
\end{equation}
where $V$ is the volume of the system.

\section{Solution of the Haldane model with staggered onsite potentials}\label{sec.S6}
In this section, we discuss several solutions of the Haldane model. First we introduce the model in real space:
\begin{equation}\label{}
    \begin{aligned}[b] 
        H=-t_0\sum_{\langle ij\rangle}\hat{a}^{\dagger}_i\hat{b}_j+t_2e^{i\phi}\sum_{\langle\langle ij\rangle\rangle}\hat{a}^{\dagger}_i\hat{a}_j+t_2e^{-i\phi}\sum_{\langle\langle ij\rangle\rangle}\hat{b}^{\dagger}_i\hat{b}_j+h.c.+\Delta\sum_i(\hat{a}^{\dagger}_i\hat{a}_j-\hat{b}^{\dagger}_i\hat{b}_j).
    \end{aligned}
\end{equation} 
Here, $ t_0 $ represents the nearest-neighbor hopping amplitude between the A and B sublattices, which are connected via three bond vectors, $ \bm{\delta}_{n} $, defined as $ \bm{\delta}_{1}=\left(\frac{1}{2},\frac{1}{2\sqrt{3}}\right) $, $ \bm{\delta}_{2}=\left(\frac{-1}{2},\frac{1}{2\sqrt{3}}\right) $, and $ \bm{\delta}_{3}=\left(0,\frac{-1}{\sqrt{3}}\right) $. The parameter $ t_2 $ corresponds to the next-nearest-neighbor hopping amplitude within the A (or B) sublattices, connected by the vectors $ \mathbf{l}_n $, where $ \mathbf{l}_1 = \left(1, 0\right) $, $ \mathbf{l}_2 = \left(\frac{-1}{2}, \frac{\sqrt{3}}{2}\right) $, and $ \mathbf{l}_3 = \left(\frac{-1}{2}, \frac{-\sqrt{3}}{2}\right) $. The phase $ \phi $ is the staggered magnetic flux, which breaks time-reversal symmetry, while $ \Delta $ refers to the onsite potential that breaks inversion symmetry.

After Fourier transformation, we have the Hamiltonian in $k$-space,
\begin{equation}
    \begin{split}
        H=\sum_{\mathbf{k}}(a^{\dagger}_{\mathbf{k}},b^{\dagger}_{\mathbf{k}})
        \begin{pmatrix} 
            \Delta+2t_2\beta^a_{\mathbf{k}}&-t_0\gamma_{\mathbf{k}}\\ -t_0\gamma^*_{\mathbf{k}}&-\Delta+2t_2\beta^b_{\mathbf{k}}
        \end{pmatrix}
        \begin{pmatrix} 
            a_{\mathbf{k}}\\ 
            b_{\mathbf{k}}
        \end{pmatrix}
    \end{split}\,,
\end{equation}
where $\gamma_{\mathbf{k}}=\sum_{n}e^{i\mathbf{k}\cdot\mathbf{\delta}_n}$, $\beta^a_{\mathbf{k}}=\sum_{n}\Re\left[e^{i\left(\mathbf{k}\cdot\mathbf{l}_n+\phi\right)}\right]=\sum_n\cos(\mathbf{k}\cdot\mathbf{l}_n+\phi)$, and $\beta^b_{\mathbf{k}}=\sum_{n}\Re\left[e^{i\left(\mathbf{k}\cdot\mathbf{l}_n-\phi\right)}\right]=\sum_n\cos(\mathbf{k}\cdot\mathbf{l}_n-\phi)$. In the main text, we set $\phi=\frac{\pi}{2}$ for the convenience. The  eigenvalues are given by the equation 
$\left(\Delta+2t_2\beta^a_{\mathbf{k}}-E\right)\left(\Delta-2t_2\beta^b_{\mathbf{k}}+E\right)+t^2_0\abs{\gamma_{\mathbf{k}}}^2=0$, which has solutions
\begin{equation}
    \begin{aligned}[b]
        \epsilon_{1,2}(\mathbf{k})=-t_2(\beta^a_{\mathbf{k}}+\beta^b_{\mathbf{k}})\pm\sqrt{t^2_0\abs{\gamma_{\mathbf{k}}}^2+t^2_2(\beta^a_{\mathbf{k}}+\beta^b_{\mathbf{k}})^2+\left(\Delta+2t_2\beta^a_{\mathbf{k}}\right)\left(\Delta-2t_2\beta^b_{\mathbf{k}}\right)}\,,
    \end{aligned}
\end{equation}
 The corresponding eigenfunctions are  
\begin{equation}
    \begin{aligned}[b]
        \ket{u_{1\mathbf{k}}}=
        \begin{pmatrix} 
            \frac{\abs{\gamma_{\mathbf{k}}}u(\mathbf{k})}{\gamma^{*}_{\mathbf{k}}\sqrt{4t^2_0\abs{\gamma_{\mathbf{k}}}^2+u^2(\mathbf{k})}}\\ 
            \frac{2t_0\abs{\gamma_{\mathbf{k}}}}{\sqrt{4t^2_0\abs{\gamma_{\mathbf{k}}}^2+u^2(\mathbf{k})}}
        \end{pmatrix},\quad
        \ket{u_{2\mathbf{k}}}=
        \begin{pmatrix} 
            \frac{\abs{\gamma_{\mathbf{k}}}v(\mathbf{k})}{\gamma^{*}_{\mathbf{k}}\sqrt{4t^2_0\abs{\gamma_{\mathbf{k}}}^2+v^2(\mathbf{k})}}\\ 
            \frac{2t_0\abs{\gamma_{\mathbf{k}}}}{\sqrt{4t^2_0\abs{\gamma_{\mathbf{k}}}^2+v^2(\mathbf{k})}}
        \end{pmatrix}.
    \end{aligned}
\end{equation}
where $-2\Delta-2t_2\beta^a_{\mathbf{k}}+2t_2\beta^b_{\mathbf{k}}+\epsilon_{2\mathbf{k}}-\epsilon_{1\mathbf{k}}=u(\mathbf{k})$, and $-2\Delta-2t_2\beta^a_{\mathbf{k}}+2t_2\beta^b_{\mathbf{k}}-\epsilon_{2\mathbf{k}}+\epsilon_{1\mathbf{k}}=v(\mathbf{k})$. 

When $t_2=0$ and $\Delta\neq0$ the model reduces to TR symmetric ``gapped graphene'', whose band dispersion is
\begin{equation}
    \begin{aligned}[b]
        \epsilon_{1,2}(\mathbf{k})=\pm\sqrt{t^2_0\abs{\gamma_{\mathbf{k}}}^2+\Delta^2}.
    \end{aligned}
\end{equation}
The above formulas are for the bulk periodic system. For the nanoribbon geometry the bands are calculated numerically, 
for TR invariant and TR non-invariant systems, both in the trivial and nontrivial topological phases. The results sre shown in Fig.~\ref{fig1}(b)-(d) of the main text. 

\section{The OMM density response in the ribbon geometry}\label{sec.S7}
This section focuses on the OMM response in the ribbon geometry. In the ribbon structure, we set $x$ as the periodic (longitudinal) direction along which the electric field is applied and $y$ as the transverse direction (with open boundary condition) along which the OMM density accumulation is computed.  The OMM density induced by the electric field is given by 
\begin{equation}\label{}
    \begin{aligned}[b]
        \delta n^{z}_{m}(y,\omega)=\frac{ieE_x}{i\eta+\omega}\int dx\sum_{n,k}\frac{\partial f_{nk}}{\partial k}n^{m}_{nn}(x,y,k)+eE_x\int dx\sum_{nn'k}\frac{f_{nk}-f_{n'k}}{\epsilon_{nk}-\epsilon_{n'k}+i\eta+\omega}n^{m}_{nn'}(x,y,k)A_{n'n}(k).
    \end{aligned}
\end{equation}
The matrix elements of the OMM density operator are given by
\begin{equation}\label{}
    \begin{aligned}[b]
        n^{m}_{nn'}(x,y,k)=\frac{1}{2}\sum_{l}[m_{nl}(k)\psi^{\dagger}_{lk}(x,y)\psi_{n'k}(x,y)+m_{ln'}(k)\psi^{\dagger}_{nk}(x,y)\psi_{lk}(x,y)].
    \end{aligned}
\end{equation}
where $ \psi_{nk}(x, y) $ (note that $\psi_{nk}(x,y)$ is not Bloch function) represents the wave function at real-space coordinates $(x, y)$ for band $n$ with wavevector $k$.  In the tight-binding basis, $\psi_{nk}(x, y)$ is expressed as
\begin{equation}\label{wave_tb_basis}
    \begin{aligned}[b]
    \psi_{nk}(x, y) = \frac{1}{\sqrt{N}} \sum_{R} \sum_{\mu} C^{nk}_{\mu} e^{ik(R + r^x_{\mu})} \phi_{R, \mu}(x, y),
    \end{aligned}
\end{equation}
where $R$ labels the periodic cells along the $ x $-direction, $ \mu $ denotes the sublattice within each periodic cell, and $ \phi_{R, \mu}(x, y) $ is the tight-binding basis function.  We assume that the latter is a $\delta$-function centered at the lattice site:
\begin{equation}\label{}
    \begin{aligned}[b]
    \phi_{R, \mu}(x, y) = \delta(x - R - r^x_{\mu}) \delta(y - r^y_{\mu}),
    \end{aligned}
\end{equation}
where $ (r^x_{\mu}, r^y_{\mu}) $ are the position coordinates of sublattice $ \mu $ in the unit cell at $ R = 0 $.  
Thus, the $x$ and the $y$-coordinates are replaced by discrete lattice-site coordinates. The coefficient $ C^{nk}_{\mu} $ in Eq.~\eqref{wave_tb_basis} can be obtained by diagonalizing the tight-binding Hamiltonian.

The integration of $x$ in function $\psi^{\dagger}_{nk}(x,y)\psi_{lk}(x,y)$ gives 
\begin{equation}\label{}
    \begin{aligned}[b]
        \int \psi^{\dagger}_{nk}(x,y)\psi_{lk}(x,y)dx=&\frac{1}{N}\sum_{RR'}\sum_{\mu\mu'}C^{nk*}_{\mu}C^{lk}_{\mu'}e^{ik(R+r^x_{\mu}-R'-r^x_{\mu'})}\delta(R+r^x_{\mu}-R'-r^x_{\mu'})\delta(y-r^y_{\mu'})\delta(y-r^y_{\mu})\\
        =&\frac{1}{N}\sum_{RR'}\sum_{\mu\mu'}C^{nk*}_{\mu}C^{lk}_{\mu'}\delta_{R,R'}\delta_{r^x_{\mu},r^x_{\mu'}}\delta_{r^y_{\mu},r^y_{\mu'}}\delta(y-r^y_{\mu})\\
        =&\sum_{\mu}C^{nk*}_{\mu}C^{lk}_{\mu}\delta(y-r^y_{\mu}).
    \end{aligned}
\end{equation}
After integrating over $x$, we obtain the matrix elements:
\begin{equation}\label{}
    \begin{aligned}[b]
        n^{m}_{nn'}(y,k)=\int n^{m}_{nn'}(x,y,k)dx=\frac{1}{2}\sum_{l\mu}[m_{nl}(k)C^{lk*}_{\mu}C^{n'k}_{\mu}+m_{ln'}(k)C^{nk*}_{\mu}C^{lk}_{\mu}]\delta(y-r^y_{\mu}).
    \end{aligned}
\end{equation}

\section{The matrix elements of the OMM operator}\label{sec.S8}
In this section, we calculate the matrix elements of the OMM operator in the ribbon geometry
\begin{equation}\label{}
    \begin{aligned}[b]
        m_{nl}(k)=&\frac{e}{4i\hbar}\mel{nk}{\{\hat{r}^x \hat{r}^y-\hat{r}^y\hat{r}^x,H\}-2\hat{r}^xH\hat{r}^y+2\hat{r}^yH\hat{r}^x}{lk}\\
        =&\frac{e}{4i\hbar}\left(\epsilon_{lk}+\epsilon_{nk}\right)\sum_{n'k'}r^x_{nn'}(k,k')r^y_{n'l}(k',k)-r^y_{nn'}(k,k')r^x_{n'l}(k',k)-\frac{e}{4i\hbar}\mel{nk}{2\hat{r}^xH\hat{r}^y-2\hat{r}^yH\hat{r}^x}{lk}\\
        =&\frac{e}{4i\hbar}\sum_{n'k'}\left(\epsilon_{lk}+\epsilon_{nk}-2\epsilon_{n'k'}\right)\left[r^x_{nn'}(k,k')r^y_{n'l}(k',k)-r^y_{nn'}(k,k')r^x_{n'l}(k',k)\right].
    \end{aligned}
\end{equation}
Employing $\hat{\mathbf{v}}=\frac{1}{i\hbar}\left[\hat{\mathbf{r}},H\right]$, and $r^x_{nn'}(k,k')=i\delta_{nn'}\partial_{k}\delta(k-k')+\delta(k-k')A_{nn'}(k)$ with $A_{nn'}(k)=i\braket{u_{nk}}{\partial_{k}u_{n'k}}$, we obtain
\begin{equation}\label{}
    \begin{aligned}[b]
        \sum_{n'k'}\left(\epsilon_{lk}+\epsilon_{nk}-2\epsilon_{n'k'}\right)r^x_{nn'}(k,k')r^y_{n'l}(k',k)&=\sum_{n'k'}\left[i\delta_{nn'}\partial_{k}\delta(k-k')+\delta(k-k')A_{nn'}(k)\right]\left(\epsilon_{lk}+\epsilon_{nk}-2\epsilon_{n'k'}\right)r^y_{n'l}(k',k)\\
        &=i\sum_{k'}\partial_{k}\delta(k-k')\left(\epsilon_{lk}+\epsilon_{nk}-2\epsilon_{nk'}\right)r^y_{nl}(k',k)+\sum_{n'}
        A_{nn'}(k)r^y_{n'l}(k)\left(\epsilon_{lk}+\epsilon_{nk}-2\epsilon_{n'k}\right)\\
        &=-2i\partial_{k}[\epsilon_{nk}]r^y_{nl}(k)+\sum_{n'}
        A_{nn'}(k)r^y_{n'l}(k)\left(\epsilon_{lk}+\epsilon_{nk}-2\epsilon_{n'k}\right)
    \end{aligned}
\end{equation}
Similarly, we have the 
\begin{equation}\label{}
    \begin{aligned}[b]
        \sum_{n'k'}\left(\epsilon_{lk}+\epsilon_{nk}-2\epsilon_{n'k'}\right)r^y_{nn'}(k,k')r^x_{n'l}(k',k)&=\sum_{n'k'}\left(\epsilon_{lk}+\epsilon_{nk}-2\epsilon_{n'k'}\right)r^y_{nn'}(k,k')\left[-i\delta_{n'l}\partial_{k}\delta(k-k')+\delta(k-k')A_{n'l}(k)\right]\\
        &=i2\partial_k[\epsilon_{lk}]r^y_{nl}(k)+\sum_{n'}\left(\epsilon_{lk}+\epsilon_{nk}-2\epsilon_{n'k}\right)r^y_{nn'}(k)A_{n'l}(k).
    \end{aligned}
\end{equation}
Finally, we have
\begin{equation}\label{}
    \begin{aligned}[b]
        m_{nl}(k)=\frac{-e}{2\hbar}\partial_{k}[\epsilon_{nk}+\epsilon_{lk}]r^y_{nl}(k)+\frac{e}{4i\hbar}\sum_{n'}\left(\epsilon_{lk}+\epsilon_{nk}-2\epsilon_{n'k}\right)\left[A_{nn'}(k)r^y_{n'l}(k)
        -r^y_{nn'}(k)A_{n'l}(k)\right].
    \end{aligned}
\end{equation}

\end{document}